\documentclass[sigplan,screen]{acmart}

\copyrightyear{2024}
\acmYear{2024}
\setcopyright{rightsretained}
\acmConference[ASPLOS '24]{29th ACM International Conference on
Architectural Support for Programming Languages and Operating Systems,
Volume 3}{April 27-May 1, 2024}{La Jolla, CA, USA}
\acmBooktitle{29th ACM International Conference on Architectural Support for
Programming Languages and Operating Systems, Volume 3 (ASPLOS '24), April
27-May 1, 2024, La Jolla, CA, USA}
\acmDOI{10.1145/3620666.3651363}
\acmISBN{979-8-4007-0386-7/24/04}

\usepackage{xfrac}  
\usepackage{textcomp} 
\usepackage{xcolor} 
\usepackage{xspace} 
\usepackage{multirow} 
\usepackage[shortlabels]{enumitem} 
\usepackage{xurl} 

\usepackage{csquotes} 
\usepackage{textgreek} 

\usepackage{subcaption} 

\usepackage{tikz} 
\usetikzlibrary{calc}
\usetikzlibrary{patterns}
\usetikzlibrary{positioning}
\usetikzlibrary{arrows}
\usetikzlibrary{arrows.meta,decorations.pathreplacing,calligraphy}
\usepackage[linesnumbered,noend,resetcount]{algorithm2e}

\usepackage{mathtools} 

\usepackage[shortcuts]{extdash} 

\usepackage[capitalize,nameinlink,noabbrev]{cleveref} 

\def\Eg{E.g.\@\xspace}
\def\eg{e.g.\@\xspace}
\def\ie{i.e.\@\xspace}
\def\cf{cf.\@\xspace}

\makeatletter
\newcommand\newtag[2]{#1\def\@currentlabel{#1}\label{#2}}
\makeatother

\newcommand{\bigland}{\bigwedge}
\newcommand{\biglor}{\bigvee}

\newcommand{\nat}{\ensuremath{\mathbb{N}}\xspace}

\newcommand{\set}[1]{\bigl\{ #1 \bigr\}}
\newcommand{\given}{\mathrel{\bigm\vert}}

\newcommand{\definedas}{\mathrel{:=}}

\providecommand{\abs}[1]{\lvert#1\rvert}

\DeclarePairedDelimiter{\floor}{\lfloor}{\rfloor}

\def\defset#1{\expandafter\def\csname#1\endcsname{\ensuremath{\mathbf{#1}}\xspace}}
\def\deffun#1{\expandafter\def\csname#1\endcsname{\ensuremath{\mathit{#1}}\xspace}}

\newcommand{\Insns}{\ensuremath{\mathbf{I}}}
\newcommand{\Ports}{\ensuremath{\mathbf{P}}}
\newcommand{\Uops}{\ensuremath{\mathbf{U}}}

\makeatletter
\def\moverlay{\mathpalette\mov@rlay}
\def\mov@rlay#1#2{\leavevmode\vtop{%
   \baselineskip\z@skip \lineskiplimit-\maxdimen
   \ialign{\hfil$\m@th#1##$\hfil\cr#2\crcr}}}
\newcommand{\charfusion}[3][\mathord]{
    #1{\ifx#1\mathop\vphantom{#2}\fi
        \mathpalette\mov@rlay{#2\cr#3}
      }
    \ifx#1\mathop\expandafter\displaylimits\fi}
\makeatother

\newcommand{\cupdot}{\charfusion[\mathbin]{\cup}{\cdot}}

\newcommand{\disjointunion}{\cupdot}

\newcommand{\inversetp}[1]{\ensuremath{\mathit{tp}^{-1}\left(#1\right)}\xspace}

\newcommand{\tpmeas}[1]{\inversetp{#1}\xspace}
\newcommand{\tpsim}[2]{\ensuremath{\mathit{tp}^{-1}_{#1}\big(#2\big)}\xspace}

\newcommand{\insnf}[1]{#1}

\newcommand{\ischeme}[1]{\texttt{#1}}

\newcommand{\opspacingindex}[2]{\,$\langle{}$\texttt{#1}$\rangle{}$}
\newcommand{\opgpr}[2]{\opspacingindex{GPR$[#1]$}{#2}}
\newcommand{\opimm}[1]{\opspacingindex{IMM$[#1]$}{R}}
\newcommand{\opmem}[2]{\opspacingindex{MEM$[#1]$}{#2}}
\newcommand{\opxmm}[1]{\opspacingindex{XMM}{#1}}
\newcommand{\opymm}[1]{\opspacingindex{YMM}{#1}}

\newcommand{\wordptrprefix}[1]{}

\newcommand{\uop}{\textmu{}op\xspace}
\newcommand{\uops}{\textmu{}ops\xspace}
\newcommand{\uopsof}[1]{\ensuremath{\mu{}\mathit{opsOf}(#1)}\xspace}

\newcommand{\ltrue}{\textit{True}\xspace}
\newcommand{\lfalse}{\textit{False}\xspace}

\newcommand{\unitgigabyte}[1]{\ensuremath{#1\,\mathrm{GB}}\xspace}

\newcommand{\pusym}{\ensuremath{\mathit{pu}}\xspace}

\newcommand{\bnsym}{\ensuremath{R_{\mathit{max}}}\xspace}

\newcommand{\relaxedalgoeps}{0.02}

\newcommand{\Utikzyposinsn}{0.0}
\newcommand{\Utikzyposuop}{-1.4}
\newcommand{\Utikzyposport}{-2.2}

\newcommand{\Utikzinsn}[3]{ \node[anchor=south] (i#1) at (#2, \Utikzyposinsn){#3}; }
\newcommand{\Utikzuop}[3]{ \node[anchor=south] (u#1) at (#2,\Utikzyposuop){#3}; }
\newcommand{\Utikzport}[3]{ \node[anchor=south] (p#1) at (#2,\Utikzyposport){#3}; }

\newcommand{\Utikzmapsitou}[4]{ \draw[] (i#1) -- node[#4 left] {#3} (u#2); }
\newcommand{\Utikzmapsutop}[2]{ \draw[] (u#1) to (p#2); }

\newcommand{\mydefaultvscale}{2.5}
\newcommand{\myvscale}{\mydefaultvscale}

\newcommand{\drawbucket}[4]{
  \draw[thick] ($(#1) + (0,\myvscale * #2)$) -- ($(#1)$) -- ($(#1) + (#3,0)$) -- ($(#1) + (#3,\myvscale * #2)$);
  \node[] at ($(#1) + (0.5 * #3,-0.4)$) {#4};
}

\newcommand{\drawinsn}[5]{
  \draw[#5, thick] let \p{A}=(#1) in
  ($(\x{A}, \myvscale * \y{A})$) rectangle ($(\x{A}, \myvscale * \y{A}) + (#3, \myvscale * #2)$) node[pos=.5] {#4};
}

\newcommand{\myCyan}{cyan!80!black}

\newcommand{\myOrange}{orange!80!white}

\newcommand{\fspace}{\quad}

\newcommand{\toolname}[1]{#1\@\xspace}

\newcommand{\pmevo}{\toolname{PMEvo}}

\newcommand{\llvmmca}{\toolname{llvm-mca}}

\newcommand{\uopsinfo}{uops.info\xspace}
\newcommand{\nanobench}{\toolname{nanoBench}}
\newcommand{\uica}{\toolname{uiCA}}
\newcommand{\facile}{\toolname{Facile}}

\newcommand{\cqa}{\toolname{CQA}}

\newcommand{\palmed}{\toolname{Palmed}}

\newcommand{\osaca}{\toolname{OSACA}}
\newcommand{\likwid}{\toolname{LIKWID}}

\newcommand{\ithemal}{\toolname{Ithemal}}

\newcommand{\granite}{\toolname{Granite}}

\SetAlgoCaptionSeparator{.}

\SetAlgorithmName{Algorithm}{Algorithm}{Algorithm}
\SetAlCapSkip{1.0em}
\SetKw{Continue}{continue}

\SetCommentSty{mycommfont}

\usepackage[]{hyperref}

\title[Explainable Port Mapping Inference with Sparse Performance Counters]{Explainable Port Mapping Inference with Sparse Performance Counters for AMD's Zen Architectures}

\author{Fabian Ritter}
\email{fabian.ritter@cs.uni-saarland.de}
\orcid{0000-0001-9227-0910}
\affiliation{%
  \institution{Saarland University, Saarland Informatics Campus}
  \city{Saarbrücken}
  \country{Germany}
}

\author{Sebastian Hack}
\email{hack@cs.uni-saarland.de}
\orcid{0000-0002-3387-2134}
\affiliation{%
  \institution{Saarland University, Saarland Informatics Campus}
  \city{Saarbrücken}
  \country{Germany}
}

\begin{document}

\begin{abstract}
\noindent
Performance models are instrumental for optimizing performance\-/sensitive code.
When modeling the use of functional units of out-of-order x86-64 CPUs, data availability varies by the manufacturer:
Instruction-to-port mappings for Intel's processors are available, whereas information for AMD's designs are lacking.
The reason for this disparity is that standard techniques to infer exact port mappings require hardware performance counters that AMD does not provide.

In this work, we modify the port mapping inference algorithm of the widely used uops.info project to not rely on Intel's performance counters.
The modifications are based on a formal port mapping model with a counter-example-guided algorithm powered by an SMT solver.
We investigate in how far AMD's processors comply with this model and where unexpected performance characteristics prevent an accurate port mapping.
Our results provide valuable insights for creators of CPU performance models as well as for software developers who want to achieve peak performance on recent AMD CPUs.

\end{abstract}

\maketitle

\section{Introduction}
\label{sec:intro}

When optimizing code for peak performance on a processor microarchitecture, understanding the architecture's performance characteristics is vital.
Modern processor designs use many techniques to improve overall performance that cause complex, irregular performance characteristics.
As manufacturers rarely provide performance models of their designs, we need to infer such models via microbenchmarks.

In this work, we consider models for an out-of-order processor's ability to exploit instruction-level parallelism: the port mapping.
The port mapping describes how instructions are decomposed into smaller operations, so-called \enquote{micro-ops} or \uops, and how these \uops are executed on the processor's execution ports.
Port mappings are important components in the cost models of compiler backends~\citep{gcc-zen-scheduling,llvm-zen-scheduling} and of instruction throughput predictors~\citep{rubial14cqa,llvmmca,llvmmca-man,laukemann18osaca,abel21uica,abel2023Facile}.

Inferring port mappings has been the subject of recent research:
\uopsinfo \citep{abel19uopsinfo} provides accurate port mappings for Intel's microarchitectures with microbenchmarks witnessing each instruction's port usage.
Their approach however does not cover other microarchitectures like AMD's recent Zen architectures, since \uopsinfo relies on Intel's per-port hardware performance counters.
Other recent port usage inference techniques have limitations that hinder their adoption in compilers and throughput predictors:
\pmevo's~\citep{ritter2020pmevo} port mappings rarely coincide with the actual microarchitecture.
In contrast to the \uopsinfo algorithm, \pmevo's evolutionary algorithm cannot provide explanatory microbenchmarks to bolster confidence in the results.
\palmed~\citep{derumigny22palmed} infers conjunctive resource mappings with good performance prediction results, but they do not map directly to the microarchitecture and therefore do not fit into existing tools.

In this paper, we present a novel port-mapping inference algorithm that does not rely on per-port performance counters and therefore supports microarchitectures that \uopsinfo cannot handle.
Our algorithm instead uses throughput measurements and a single hardware performance counter to count the total number of executed \uops for a given microbenchmark.
Assuming that the processor behaves according to a formal port mapping model, these measurements are sufficient to infer port mappings with explanatory microbenchmarks similar to the \uopsinfo algorithm.
The core of our technique is a counter-example-guided algorithm powered by a satisfiability-modulo-theories (SMT) solver that directly leverages the formal model.

We evaluate our approach in a case study on AMD's Zen+ architecture.
We report cases where irregular performance characteristics and undocumented or wrongly documented cases hinder an accurate performance model of Zen+.
Using our algorithm, we infer the most comprehensive explainable port mapping for Zen+ to date.
Our port mapping outperforms the state of the art in terms of throughput prediction accuracy for the supported instructions.

In summary, we provide the following contributions:
\begin{itemize}[leftmargin=*]
  \item An explainable inference algorithm for port mappings that does not require Intel's per-port performance counters and
  \item an implementation that we evaluate on AMD's Zen+ architecture, which was previously out of scope for explainable port mapping inference algorithms.
  \item The result is, to the best of our knowledge, the most comprehensive and accurate port mapping available for Zen+.
  \item Our case study documents numerous previously undocumented or misdocumented aspects of Zen+.
\end{itemize}

\section{Background}
\label{sec:background}

\subsection{Out-of-Order Microarchitectures}

Modern microarchitectures are complex designs that combine various techniques to improve performance.
Typically, they decode instructions into one or more \uops and apply out-of-order execution:
The \uops are reordered and executed in parallel on the processor's functional units as long as their read-after-write dependencies are preserved~\citep{tomasulo67,hennessy17}.

\begin{figure}
  \centering
  \scalebox{.95}{
    \begin{tikzpicture}

      \draw[] (-3,-1.0) rectangle (-0.8,-1.6);
      \node[anchor=north] at (-1.9,-1.05) (uopcache) {\uop Cache};

      \draw[-latex] (-1.9, -1.6) -- (-1.9, -1.9);

      \draw[] (-0.4,-1.0) rectangle (3,-1.6);
      \node[anchor=north] at (1.3,-1.05) (decode) {Instruction Decoder};

      \draw[-latex] (1.3, -1.6) -- (1.3, -1.9);
      \draw[-latex] (-0.4, -1.3) -- (-0.8, -1.3);

      \draw[] (-3,-1.9) rectangle (3,-2.5);
      \node[] at (0,-2.2) (reg) {Register Management};

      \draw[-latex] (0.0, -2.5) -- (0.0, -2.8);

      \draw[] (-3,-2.8) rectangle (3,-3.4);
      \node[] at (0,-3.1) (sched) {Scheduler};

      \draw[-latex] (-2.325, -3.4) -- (-2.325, -3.7);
      \draw[-latex] (-0.775, -3.4) -- (-0.775, -3.7);
      \draw[-latex] (0.775, -3.4) -- (0.775, -3.7);
      \draw[-latex] (2.325, -3.4) -- (2.325, -3.7);

      \draw[] (-3,-3.7) rectangle (-1.65,-4.3);
      \node[] at (-2.325,-4.0) (p0) {Port 0};

      \draw[] (-3,-4.3) rectangle (-1.65,-5.6);

      \node[anchor=north, align=center] at (-2.325,-4.3) {\small Int ALU\\\small Vec ALU\\\small DIV};

      \draw[] (-1.45,-3.7) rectangle (-0.1,-4.3);
      \node[] at (-0.775,-4.0) (p1) {Port 1};

      \draw[] (-1.45,-4.3) rectangle (-0.1,-5.6);

      \node[anchor=north, align=center] at (-0.775,-4.3) {\small Int ALU\\\small Vec ALU};

      \draw[] (0.1,-3.7) rectangle (1.45,-4.3);
      \node[] at (0.775,-4.0) (p2) {Port 2};

      \draw[] (1.45,-4.3) rectangle (0.1,-5.6);

      \node[anchor=north, align=center] at (0.775,-4.3) {\small Memory\\Load \&\\Store};

      \draw[] (1.65,-3.7) rectangle (3,-4.3);
      \node[] at (2.325,-4.0) (p3) {Port 3};

      \draw[] (3,-4.3) rectangle (1.65,-5.6);

      \node[anchor=north, align=center] at (2.325,-4.3) {\small Memory\\Load};

      \draw[] (3.6,-5.6) rectangle (4.8,-3.4);
      \node[rotate=90] at (4.2,-4.5) (dcache) {L1 DCache};
      \draw[latex-] (4.2, -5.6) -- (4.2, -5.85) -- (2.325, -5.85) -- (2.325, -5.6);
      \draw[] (2.325, -5.85) -- (0.775, -5.85) -- (0.775, -5.6);

      \draw[] (3.6,-1.0) rectangle (4.8,-3.2);
      \node[rotate=90] at (4.2,-2.1) (dcache) {L1 ICache};
      \draw[-latex] (3.6, -1.3) -- (3.0, -1.3);
    \end{tikzpicture}
  }
  \caption{
    Simplified overview of a modern processor design (based on Figure 2-8 in the Intel Software Optimization Manual~\citep[Section 2.6]{intel-opt-manual}).
  }
  \label{fig:proc_schematic}
  \Description{Diagram showing the flow of information between components of a modern processor core.
    Instructions are loaded from memory via the Instruction Cache.
    The Instruction Decoder decodes them into micro-ops, which flow to a micro-op cache.
    The Register Management unit obtains micro-ops from the micro-op cache or directly from the Instruction Decoder.
    From Register Management, the micro-ops go to a Scheduler, which assigns them among four Ports.
    Port 0 handles integer and vector arithmetic, as well as division.
    Port 1 handles integer and vector arithmetic.
    Port 2 handles memory loads and stores.
    Port 3 handles memory loads.
    The Ports that handle memory loads and stores also connect to the Data Cache.
  }
\end{figure}
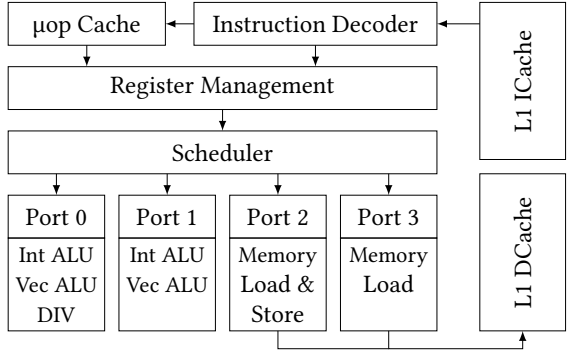

\cref{fig:proc_schematic} illustrates such a microarchitecture.
Instructions are loaded from memory via the instruction cache.
They are decoded into \uops, which are cached for future re-use.
Their register operands are translated to a larger set of microarchitectural registers.
This translation eliminates \enquote{false} write-after-read and write-after-write dependencies that would otherwise limit reordering.
Finally, a scheduler assigns each \uop, once its operands are ready, to a port with the appropriate functional unit.
Most functional units of modern processors are pipelined, allowing the ports to accept one \uop per cycle.
To understand how an instruction is executed, we need to know its \uop decomposition and which ports can execute these \uops.
This information is the \emph{port mapping}.

Processors often provide hardware performance counters: additional registers to count events in the processor.
They gather statistics on how the processor executes code without affecting the execution itself.
Which events can be counted depends on the microarchitecture, \eg, executed instructions and \uops or cache misses.
The port mapping inference algorithm of \uopsinfo~\citep{abel19uopsinfo} depends on counting the \uops that each individual port executes.
Relevant microarchitectures, \eg, AMD's Zen family, do not provide these counters.
Tools like \nanobench \citep{abel20nanobench} and \likwid \citep{treibig10likwid,gruber22likwid} can read performance counters and measure the throughput for microbenchmarks.

\subsection{The Port Mapping Model}
\label{sec:background:pm_model}

This work builds on a formal port mapping model introduced in previous work~\citep{abel19uopsinfo,ritter2020pmevo}:
Port mappings are tripartite graphs $M\definedas(\Insns\disjointunion\Uops\disjointunion\Ports, F\disjointunion E)$ between a set $\Insns$ of instruction schemes, a set $\Uops$ of \uops, and a set $\Ports$ of execution ports.
Instruction schemes (or instruction forms) abstract sets of concrete instructions that differ in their operands.
The x86-64 instruction set contains thousands of instruction schemes.
\Eg, the following scheme represents a 64-bit addition with two general purpose register operands:
\begin{center}
  \ischeme{add\opgpr{64}{RW},\opgpr{64}{R}}
\end{center}

In a port mapping, the labeled edges $F\subseteq\Insns\times\nat\times\Uops$ describe how many \uops of a specific kind the instruction schemes require.
The unlabeled edges $E\subseteq \Uops\times\Ports$ between \uops and ports capture where each \uop can be executed.

\begin{figure}
  \centering
  \subfloat[]{
    \label{fig:ex_uopsinfo_algo:mapping}
    \scalebox{0.9}{
    \begin{tikzpicture}
      \node[anchor=south] (icap) at (-2.2, \Utikzyposinsn) {\Insns:};
      \node[anchor=south] (ucap) at (-2.2, \Utikzyposuop) {\Uops:};
      \node[anchor=south] (pcap) at (-2.2, \Utikzyposport) {\Ports:};
      \Utikzinsn{1}{-1.5}{\ischeme{add}}
      \Utikzinsn{2}{0.0}{\ischeme{fma}}
      \Utikzinsn{3}{1.5}{\ischeme{mul}}

      \Utikzuop{1}{-0.75}{$u_1$}
      \Utikzuop{2}{0.75}{$u_2$}

      \Utikzport{1}{-0.75}{$p_1$}
      \Utikzport{2}{0.75}{$p_2$}

      \Utikzmapsitou{1}{1}{1}{}
      \Utikzmapsitou{2}{1}{2}{}
      \Utikzmapsitou{2}{2}{1}{}
      \Utikzmapsitou{3}{2}{1}{}
      \Utikzmapsutop{1}{1}
      \Utikzmapsutop{1}{2}
      \Utikzmapsutop{2}{2}
    \end{tikzpicture}
    }
  }
  \hspace*{-0.9cm}
  \subfloat[]{
    \label{fig:ex_uopsinfo_algo:basic_example}
    \newcommand{\scaleboxfactor}{0.85}
    \newcommand{\scaletikzfactor}{0.7}
    \newcommand{\bucketwidth}{2.3}
    \newcommand{\bucketheight}{1.8}
    \newcommand{\bucketpositionA}{0.0}
    \newcommand{\bucketpositionB}{2.8}
    \newcommand{\bucketpositionC}{5.6}
    \renewcommand{\myvscale}{1.8}
    \scalebox{\scaleboxfactor}{
      \begin{tikzpicture}[scale=\scaletikzfactor]
        \drawbucket{\bucketpositionA,0}{\bucketheight}{\bucketwidth}{$p_1$}
        \drawbucket{\bucketpositionB,0}{\bucketheight}{\bucketwidth}{$p_2$}

        \drawinsn{\bucketpositionA,0}{1.0}{\bucketwidth}{$u_1$ (\ischeme{fma})}{pattern color=\myOrange,pattern=north west lines}
        \drawinsn{\bucketpositionB,0}{1.0}{\bucketwidth}{$u_2$ (\ischeme{mul})}{pattern color=\myCyan,pattern=crosshatch}
        \drawinsn{\bucketpositionB,1.0}{0.5}{\bucketwidth}{$u_2$ (\ischeme{fma})}{pattern color=\myOrange,pattern=north west lines}

        \draw[->, thick] (-0.4, 0.0) -- ($(-0.4, \bucketheight*\myvscale)$);
        \foreach \y in {0, 1, 2, 3}%
            \draw[thick] ($(-0.4, 0.5*\y*\myvscale) + (0.1, 0.0)$) -- ($(-0.4, 0.5*\y*\myvscale) + (-0.1, 0.0)$) node[left] {$\y$};

        \draw[thick] ($(-0.5, \myvscale * 1.5)$) -- ($(5.5, \myvscale * 1.5)$);
        \node[anchor=west] at ($(5.5, \myvscale* 1.5)$) {$t$};

      \end{tikzpicture}
    }
    \renewcommand{\myvscale}{\mydefaultvscale}
  }
  \caption{Example port mapping (a) and optimal \uop distribution for [\ischeme{mul}, \ischeme{mul}, \ischeme{fma}] (b).
    The processor executes two~$u_1$ \uops (for the \ischeme{fma} instruction) and three~$u_2$ \uops (one for the \ischeme{fma} instruction and one for each \ischeme{mul} instruction) for this instruction sequence.
    Only port $p_2$ can handle $u_2$ while $u_1$ could be executed on either port. }
  \Description{The port mapping (a) relates three instructions (add, fma, and mul) with two micro-ops (u1 and u2) and two ports (p1 and p2).
    add is decomposed into one u1 micro-op.
    fma is decomposed into two u1 micro-ops and one u2 micro-op.
    mul is decomposed into one u2 micro-op.
    The u1 micro-ops can be executed on either port p1 or p2.
    The u2 micro-ops can only be executed on port p2.
    The optimal uop distribution (b) illustrates a steady state execution of the instruction sequence [mul, mul, fma].
    Port p1 is occupied for two cycles by the two u1 micro-ops of the fma instruction.
    Port p2 is occupied for three cycles by the two u2 micro-ops of the mul instructions and the u2 micro-op of the fma instruction.
    The objective value t, i.e., the inverse throughput, is the greater of the two: three cycles.
  }
\end{figure}
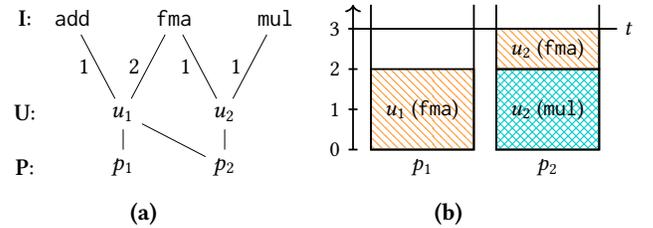
For an example, consider the port mapping in \cref{fig:ex_uopsinfo_algo:mapping} for the instruction schemes \ischeme{add}, \ischeme{mul}, and \ischeme{fma} (\enquote{fused multiply and add}) of a simplified microarchitecture.
\ischeme{add} and \ischeme{mul} are each decomposed into a single \uop: $u_1$ and $u_2$.
$u_2$ has one execution port whereas $u_1$ can be executed on either $p_1$ or $p_2$.
Executing the \ischeme{fma} instruction requires processing two $u_1$ \uops and one $u_2$ \uop.

The port mapping model links a processor's port mapping and the throughput of dependency-free instruction sequences~$e$ via the following linear program (LP):
\begin{alignat*}{3}
  &\text{min.}& \quad & t \\
  &\text{s.t.}& \quad &\displaystyle\sum_{k\in\Ports} x_{uk} = \hspace{-0.6em}\displaystyle\sum_{(i, n, u)\in F} \hspace{-0.8em}\insnf{e}(i) \cdot n && \fspace\text{for all \uops $u\in \Uops$}\tag{A}\label{js3A}\\
  &&&  \displaystyle\sum_{u\in\Uops} x_{uk} = p_k      && \fspace\text{for all ports $k\in\Ports$}\tag{B}\label{js3B}\\
  &&&  p_k \le t                             && \fspace\text{for all ports $k\in\Ports$}\tag{C}\label{js3C}\\
  &&&  x_{uk} \ge 0                          && \fspace\begin{gathered}\text{for all \uops $u\in\Uops$,}\\\text{ports $k\in\Ports$}\end{gathered}\tag{D}\label{js3D}\\
  &&&  x_{uk} = 0                            && \fspace\text{if $ (u, k) \not\in E$}\tag{E}\label{js3E}
\end{alignat*}

The optimal objective value of this linear program is the inverse throughput $\tpsim{M}{e}$ of~$e$ with the port mapping~$M$, \ie, the average number of processor cycles required for an instance of $e$ when it is executed indefinitely in a steady state.

Intuitively, each \uop of the executed instructions contributes a share of \enquote{mass} corresponding to one cycle of utilization of a port.
The non-negative real-valued $x_{uk}$ variables represent the mass contributed by $u$~\uops that is executed on port~$k$.
$\insnf{e}(i)$ denotes how often the instruction scheme~$i$ occurs in~$e$.
Constraint~\ref{js3A} therefore ensures that all mass is distributed to the $x_{uk}$ variables.
Constraint~\ref{js3B} establishes real-valued $p_k$~variables for the total mass assigned to each port~$k$ and constraint \ref{js3C} introduces~$t$ as an upper bound to the total masses of every port.
Lastly, constraint \ref{js3E} ensures that \uops are only assigned to compatible ports.
The requirement to minimize the upper bound~$t$ guarantees that the \uop mass is distributed as evenly as possible with the port mapping, achieving peak throughput.
Solutions that assign non-integer \uop masses to ports correspond to steady-state executions where \uops go to different ports in different repetitions of the instruction sequence.
This model assumes that the port mapping is the only source of throughput bottlenecks.

\cref{fig:ex_uopsinfo_algo:basic_example} visualizes an optimal LP solution for the port mapping from \cref{fig:ex_uopsinfo_algo:mapping} and the instruction sequence [\ischeme{mul}, \ischeme{mul}, \ischeme{fma}].
The mass for each port is collected in a corresponding bucket.
The objective value~$t$ is the mass of the highest bucket: three cycles.
This mass is caused by the two $u_2$ \uops of the \ischeme{mul} instructions and the $u_2$ \uop of the \ischeme{fma} instruction.
While the $u_1$ \uops could also be executed on $p_2$, they need to be handled by port $p_1$ for an optimal distribution.

\subsection{The uops.info Algorithm}

Our work follows prior approaches \citep{abel19uopsinfo,ritter2020pmevo} with the goal to infer a processor's port mapping from microbenchmarks.
We base our approach on a port mapping inference algorithm by Abel and Reineke~\citep[Section 5.1]{abel19uopsinfo}, which relies on blocking instructions:
An instruction blocks a port subset $\pusym$ if it executes as a single \uop that can use exactly the ports in $\pusym$.
The algorithm needs blocking instructions for the port sets of every \uop.
Abel and Reineke find blocking instructions by executing each instruction scheme in a steady state while counting the \uops executed per port via performance counters.
If there are as many \uops as instructions, $i$~is a blocking instruction for the ports where its \uops can be observed.

For an example, consider a 32-bit integer addition with register operands on the Intel Skylake microarchitecture:
\begin{center}
  \ischeme{add\opgpr{32}{RW},\opgpr{32}{R}}
\end{center}
When we benchmark this instruction scheme in a steady state, Skylake's performance counters show that
\begin{enumerate}[leftmargin=*]
  \item an \ischeme{add} instruction requires 0.25 cycles on average, \ie, four of them can execute in a single cycle,
  \item the processor executes one \uop per \ischeme{add} instruction, and
  \item the \uops are executed in equal parts on ports 0, 1, 5, and 6 of the microarchitecture (which has a total of 8 ports).
\end{enumerate}
This \ischeme{add} instruction scheme therefore blocks the port set $\set{0, 1, 5, 6}$.
Observation~3 requires per-port \uop counters that are not available in AMD's processors.

\begin{algorithm}
  \normalsize
  \DontPrintSemicolon
  \KwIn{instruction under investigation $i$}
  $\mathit{blkInsns} \leftarrow \text{pairs of (blocking insn, blocked ports),}$ \hspace*{0.8cm}$\text{sorted by ascending number of blocked ports}$\;
  $\mathit{foundUops} \leftarrow \{\}$\;\label{relaxed:algo:uopsinfo:def_foundUops}

  \For{$(B, \pusym) \in \mathit{blkInsns}$}{
    $k \leftarrow \text{\# blocking insns $B$ sufficient to flood $\pusym$}$\;\label{relaxed:algo:uopsinfo:def_k}
    $\mathit{uops} \leftarrow \mathit{measureUopsOnPorts}([\,k\times B, i\,], \pusym)$\;\label{relaxed:algo:uopsinfo:run_exp}
    $\mathit{surplusUops} \leftarrow \mathit{uops} - k$\;

    \For{$\pusym', n \in \mathit{foundUops}$}{\label{relaxed:algo:uopsinfo:subloop_start}
      \If{$\pusym' \subset \pusym$}{
        $\mathit{surplusUops} \leftarrow  \mathit{surplusUops} - n$\;
      }
    }\label{relaxed:algo:uopsinfo:subloop_end}
    \If{$\mathit{surplusUops} > 0$}{
      $\mathit{foundUops}[\pusym] \leftarrow \mathit{surplusUops}$\;\label{relaxed:algo:uopsinfo:res}
    }
  }
  \Return{$\mathit{foundUops}$}

  \caption{Port mapping inference for \uopsinfo~\citep{abel19uopsinfo}.}
  \label{relaxed:algo:uopsinfo}
\end{algorithm}

Abel and Reineke select one blocking instruction for each occurring port set and apply \cref{relaxed:algo:uopsinfo} for each instruction scheme~$i$.
The multiset $\mathit{foundUops}$ of \uops, which are represented by their sets of admissible ports, for~$i$ is filled throughout a run of the algorithm.
The algorithm benchmarks~$i$ with each blocking instruction~$B$ for a set~$\pusym$ of ports, starting with the smallest port sets and proceeding to increasing port set sizes.
Each microbenchmark contains~$i$ and enough copies of the considered blocking instruction~$B$ such that any \uop that \emph{can} be executed on ports outside of~$\pusym$ \emph{is} executed on these alternative ports (ll.\ \ref{relaxed:algo:uopsinfo:def_k}, \ref{relaxed:algo:uopsinfo:run_exp}).
The number~$k$ of blocking instruction copies must ensure that each blocked port in~$\pusym$ receives at least as many \uops as~$i$ uses:
\begin{equation}
  k \geq \abs{\pusym} \cdot \uopsof{i}
  \label{relaxed:k_condition}
\end{equation}
Otherwise, ports in $\pusym$ might be unoccupied while \uops of~$i$ are issued, allowing \uops of $i$ on $\pusym$ even though they could be executed on other ports.

When running this microbenchmark, Abel and Reineke count executed \uops on ports in~$\pusym$ via per-port performance counters.
The result is the sum of $k$ and the number of \uops of~$i$ that only use ports in~$\pusym$.
These surplus \uops include not only \uops that use \emph{all} ports of~$\pusym$ but also those that have only a subset of~$\pusym$ available.
Because we assume a blocking instruction for the port set of each occurring \uop and because the blocking instructions are sorted by ascending number of blocked ports, all \uops for proper subsets of~$\pusym$ were characterized in previous iterations of the loop.
We can therefore subtract these previously characterized \uops from the surplus \uops (ll.\ \ref{relaxed:algo:uopsinfo:subloop_start}--\ref{relaxed:algo:uopsinfo:subloop_end}) to obtain the \uops of~$i$ that can use any port in~$\pusym$ (l.\ \ref{relaxed:algo:uopsinfo:res}).
The port usage of~$i$ is fully characterized once every blocking instruction has been considered.

For example, consider a processor with the port mapping in \cref{fig:ex_uopsinfo_algo:mapping}.
There are two blocking instructions: \ischeme{mul} for the port set $\set{p_2}$ and \ischeme{add} for $\set{p_1, p_2}$.
For this example, to characterize the \ischeme{fma} instruction, we use $k \definedas \abs{\pusym} \cdot \uopsof{i} = \abs{\pusym} \cdot 3$ blocking instructions per benchmark.

\begin{figure}
  \newcommand{\scaleboxfactor}{0.82}
  \newcommand{\scaletikzfactor}{0.65}
  \newcommand{\bucketwidth}{2.4}
  \newcommand{\bucketheight}{2.35}
  \newcommand{\bucketpositionA}{0.0}
  \newcommand{\bucketpositionB}{2.8}
    \renewcommand{\myvscale}{1.9}
  \centering
  \subfloat[]{
    \label{fig:ex_uopsinfo_algo:with_mul}
    \scalebox{\scaleboxfactor}{
      \begin{tikzpicture}[scale=\scaletikzfactor]
        \drawbucket{\bucketpositionA,0}{\bucketheight}{\bucketwidth}{$p_1$}
        \drawbucket{\bucketpositionB,0}{\bucketheight}{\bucketwidth}{$p_2$}

        \drawinsn{\bucketpositionA,0}{1.0}{\bucketwidth}{$u_1$ (\ischeme{fma})}{pattern color=\myOrange,pattern=north west lines}
        \drawinsn{\bucketpositionB,0}{1.5}{\bucketwidth}{$u_2$ (\ischeme{mul})}{pattern color=\myCyan,pattern=crosshatch}
        \drawinsn{\bucketpositionB,1.5}{0.5}{\bucketwidth}{$u_2$ (\ischeme{fma})}{pattern color=\myOrange,pattern=north west lines}

        \draw[->, thick] (-0.4, 0.0) -- ($(-0.4, \bucketheight*\myvscale)$);
        \foreach \y in {0, 1, 2, 3, 4}%
            \draw[thick] ($(-0.4, 0.5*\y*\myvscale) + (0.1, 0.0)$) -- ($(-0.4, 0.5*\y*\myvscale) + (-0.1, 0.0)$) node[left] {$\y$};

      \end{tikzpicture}
    }
  }
  \subfloat[]{
    \label{fig:ex_uopsinfo_algo:with_add}
    \scalebox{\scaleboxfactor}{
      \begin{tikzpicture}[scale=\scaletikzfactor]
        \drawbucket{\bucketpositionA,0}{\bucketheight}{\bucketwidth}{$p_1$}
        \drawbucket{\bucketpositionB,0}{\bucketheight}{\bucketwidth}{$p_2$}

        \drawinsn{\bucketpositionA,0}{1.5}{\bucketwidth}{$u_1$ (\ischeme{add})}{pattern color=\myCyan,pattern=crosshatch}
        \drawinsn{\bucketpositionB,0}{1.5}{\bucketwidth}{$u_1$ (\ischeme{add})}{pattern color=\myCyan,pattern=crosshatch}

        \drawinsn{\bucketpositionA,1.5}{0.75}{\bucketwidth}{$u_1$ (\ischeme{fma})}{pattern color=\myOrange,pattern=north west lines}
        \drawinsn{\bucketpositionB,1.5}{0.5}{\bucketwidth}{$u_2$ (\ischeme{fma})}{pattern color=\myOrange,pattern=north west lines}
        \drawinsn{\bucketpositionB,2.0}{0.25}{\bucketwidth}{$u_1$ (\ischeme{fma})}{pattern color=\myOrange,pattern=north west lines}

        \draw[->, thick] (-0.4, 0.0) -- ($(-0.4, \bucketheight*\myvscale)$);
        \foreach \y in {0, 1, 2, 3, 4}%
            \draw[thick] ($(-0.4, 0.5*\y*\myvscale) + (0.1, 0.0)$) -- ($(-0.4, 0.5*\y*\myvscale) + (-0.1, 0.0)$) node[left] {$\y$};
      \end{tikzpicture}
    }
  }

  \caption{
    Possible steady-state distributions of \uops per port in benchmarks of \ischeme{fma} with (a) 3 \ischeme{mul} and (b) 6 \ischeme{add} blocking instructions, using the port mapping from \cref{fig:ex_uopsinfo_algo:mapping}.
  }
  \label{fig:ex_uopsinfo_algo}
  \Description{
    Each subfigure shows a steady-state distribution of micro-ops to the ports p1 and p2.
    In (a), the fma instruction is benchmarked with 3 mul blocking instructions.
    The u2 micro-ops of the mul blocking instructions utilize p2 for three cycles.
    The u1 micro-ops of the fma instruction utilize p1 for two cycles.
    The remaining u2 micro-op of the fma instruction utilizes p2 for an additional cycle.
    The inverse throughput is the total utilization of p2: 4 cycles.
    In (b), the fma instruction is benchmarked with 6 add blocking instructions.
    The 6 u1 micro-ops of the add blocking instructions utilize both p1 and p2 for 3 cycles.
    The u2 micro-op of the fma instruction adds 1 cycle of utilization to p2.
    The 2 u1 micro-ops of the fma instruction utilize p1 for 1.5 cycles and p2 for 0.5 cycles.
    Both ports are therefore utilized for 4.5 cycles, which is the resulting inverse throughput.
  }
\end{figure}
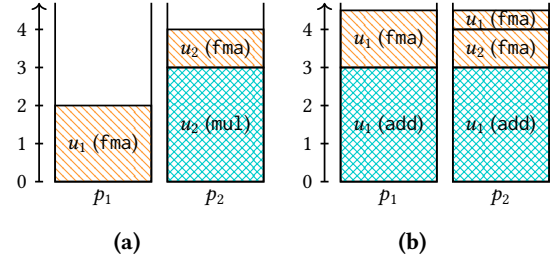

The algorithm first benchmarks \ischeme{fma} with blocking instructions for port sets of size~1, \ie, \ischeme{mul}, with $k = 3$.
\cref{fig:ex_uopsinfo_algo:with_mul} shows the distribution of \uops in a steady state execution:
Four \uops execute on the blocked port set $\set{p_2}$.
Since no \uops were characterized for smaller port sets, we conclude that \ischeme{fma} uses $4-k=1$ \uop that can be executed on $\set{p_2}$.

Next, \ischeme{fma} is benchmarked with the \ischeme{add} instruction, which blocks a port set of size~2 (\ie, $k = 6$).
As shown in \cref{fig:ex_uopsinfo_algo:with_add}, we count nine \uops on the ports $\set{p_1, p_2}$.
Subtracting six blocking instructions, three surplus \uops remain.
One of these is explained by the $\set{p_2}$ \uop found previously.
The two remaining \uops have the entire port set $\set{p_1, p_2}$ available.
With no more blocking instructions remaining, we obtain the port usage for \ischeme{fma}:
\(\set{2\times \set{p_1, p_2}, 1\times \set{p_2}}\)

In practice, the \uopsinfo implementation\footnote{\url{https://github.com/andreas-abel/nanoBench/blob/faf75236cade57f7927f9ee949ebc679fc7864b7/tools/cpuBench/cpuBench.py\#L3393}} computes the number~$k$ of blocking instructions (l.\ \ref{relaxed:algo:uopsinfo:def_k}) as follows:
\[
  \begin{multlined}k \leftarrow \min \bigl(100, \max\bigl(10, \abs{\pusym} \cdot \uopsof{i}, \\ 2 \cdot \abs{\pusym} \cdot \max(1,\floor{\tpmeas{[\,i\,]}})\bigr)\bigr)\end{multlined}
\]
The resulting $k\in[10, 100]$ depends on the number~$\abs{\pusym}$ of blocked ports, the cycles~$\tpmeas{[\,i\,]}$ required to execute~$i$ in a steady state, and $i$'s number $\uopsof{i}$ of \uops.
This term fulfills constraint~\ref{relaxed:k_condition} for reasonable numbers of \uops.
Compared to an implementation that satifisies constraint~\ref{relaxed:k_condition} tightly, we expect more resilience against measurement errors from the larger number of blocking instructions.

A key benefit of this port mapping inference algorithm is that the performed microbenchmarks serve as witnesses for the result:
For each instruction~$i$ and each port set~$\pusym$, there is an experiment explaining if~$i$ uses a \uop for~$\pusym$.

\section{Inferring Port Mappings Without Per-Port \uop Counters}
\label{sec:main}

For AMD's Zen microarchitectures and several ARM designs, the \uopsinfo algorithm is not applicable as they lack performance counters for executed \uops per port.
The key insight of our work is that the problems requiring per-port \uop counters in the \uopsinfo algorithm can be solved without them if we assume that the processor follows the port mapping model for some (unknown) port mapping.
In the following, we provide alternatives that do not use per-port \uop counters for the relevant parts of the \uopsinfo algorithm.
The only performance counter used, besides time measurements, counts the total number of \uops executed for a microbenchmark.

\subsection{Counting \uops that Cannot Avoid Blocked Ports}
\label{relaxed_algo:ssec:evading_insns}

To determine how many \uops run on ports that are flooded with blocking instructions, consider two experiments:
\begin{itemize}[leftmargin=*]
  \item
$e\definedas [\,k\times B, i\,]$ consist of an instruction~$i$ under investigation and $k$~blocking instructions~$B$ for a port set~$\pusym$.
 \item
$e' \definedas [\,k\times B\,]$ contains only the blocking instructions.
\end{itemize}
If all \uops of~$i$ can use unblocked ports, $\tpmeas{e} = \tpmeas{e'}$ holds.
Otherwise, each \uop of~$i$ that needs a port in $\pusym$ utilizes a flooded port for one cycle per iteration.
Each such \uop therefore adds $\sfrac{1}{\abs{\pusym}}$ to the observed inverse throughput, \ie:
\begin{align*}
  &\tpmeas{e} = \tpmeas{e'} + \sfrac{1}{\abs{\pusym}} \cdot \text{\uops of $i$ executed on $\mathit{pu}$}\\
  \Leftrightarrow&\ \text{\uops of $i$ executed on $\mathit{pu}$} = \big(\tpmeas{e} - \tpmeas{e'}\big) \cdot \abs{\pusym}
\end{align*}

For example, reconsider \cref{fig:ex_uopsinfo_algo:with_mul}.
The $u_2$ \uop of the \ischeme{fma} instruction cannot evade from the blocked port $p_2$.
Compared to an execution of only the blocking instructions (3 cycles), the inverse throughput is increased by $\sfrac{1}{\abs{\pusym}} = 1$ cycle.

\subsection{Identifying Unique Blocking Instructions}
\label{relaxed_algo:ssec:find_blocking_insns}

We find and characterize blocking instructions as follows:
\begin{enumerate}[leftmargin=*]
  \item Count the \uops when executing each instruction individually.
    Each instruction with only a single \uop is a candidate.

  \item Determine for each candidate~$i$ the number of ports on which its \uop can be executed, \ie, the number of instances of~$i$ that can be executed per cycle.
    We measure this as the (non-inverse) throughput of~$i$: $1/\tpmeas{[\,i\,]}$.

  \item Filter redundant candidates, leaving one blocking instruction per port set.
    Two candidates cannot be redundant if their port sets have different sizes.
    For two candidates~$i$ and~$j$ with equally sized port sets, we check for redundancy by measuring if their inverse throughputs are additive:
    The port sets of~$i$ and~$j$ are equal if
    \[\tpmeas{[\,i, j\,]} = \tpmeas{[\,i\,]} + \tpmeas{[\,j\,]}\]

  \item Infer the port mapping of the remaining blocking instructions.
    Compared to the full inference problem, this concerns only few instruction schemes -- \eg, the \uopsinfo port mapping for Intel's Skylake has 12 distinct port sets that require blocking instructions -- and every scheme uses only a single \uop with a known number of ports.
    This makes the computationally expensive algorithm described in the following section practical.
\end{enumerate}

\subsection{Counter-Example Guided Port Mapping Inference}
\label{relaxed_algo:ssec:cegpmi}

The port mapping model relates the throughput achieved for given instruction sequences with the processor's port mapping.
We exploit this relation with inspiration from counter-example-guided abstraction refinement~\citep{clarke00cegar} and counter-example-guided inductive synthesis~\citep{solarlezama06cegis}.

\begin{algorithm}
  \DontPrintSemicolon
    $\mathit{Exps} \gets \set{}$\;
    \While{$\mathsf{true}$}{
      $m1 \gets \mathit{findMapping}(\mathit{Exps})$\;\label{algo:cegpminf:line_find}
      \lIf{$m1 = \mathsf{None}$}{
        \Return $\mathsf{None}$\label{algo:cegpminf:line_fail}
      }
      $m2, \mathit{newExp} \gets \mathit{findOtherMapping}(\mathit{Exps}, m1)$\;\label{algo:cegpminf:line_findother}
      \lIf{$m2 = \mathsf{None}$}{
        \Return m1\label{algo:cegpminf:line_success}
      }
      $\mathit{cycles} \gets \mathit{measureCycles}(\mathit{newExp})$\;\label{algo:cegpminf:line_measure}
      $\mathit{Exps} \gets \mathit{Exps} \cup \{(\mathit{newExp}, \mathit{cycles})\}$\;\label{algo:cegpminf:line_extend}
    }
  \caption{Counter-example-guided inference.}
  \label{algo:cegpminf}
\end{algorithm}

\cref{algo:cegpminf} shows the high-level structure of our counter-example-guided port mapping inference algorithm.
It is centered around a set $\mathit{Exps}$ of microbenchmarks annotated with the inverse throughput measured on the processor under investigation.
In each iteration, we search a port mapping~$m1$ that leads to the measured inverse throughputs in $\mathit{Exps}$~(l.\,\ref{algo:cegpminf:line_find}).
If no mapping is found, the observations do not match the model: the algorithm terminates unsuccessfully~(l.\,\ref{algo:cegpminf:line_fail}).
Otherwise, we search for a different port mapping~$m2$ that is also consistent with the measurements in $\mathit{Exps}$, but that is distinguished from~$m1$ by an experiment~$\mathit{newExp}$~(l.\,\ref{algo:cegpminf:line_findother}).
This means that~$m1$ and~$m2$ yield the same throughputs for $\mathit{Exps}$, but different throughputs for~$\mathit{newExp}$.
If no such mapping and experiment exist, $m1$ is indistinguishable by throughput measurements from the processor's actual port mapping~(l.\,\ref{algo:cegpminf:line_success}).
Otherwise, we measure the inverse throughput of~$\mathit{newExp}$, add it to $\mathit{Exps}$ (ll.\,\ref{algo:cegpminf:line_measure}-\ref{algo:cegpminf:line_extend}), and continue with the next iteration.

\begin{figure}[tp]
  \centering
  \subfloat[]{
    \begin{tikzpicture}
      \renewcommand{\Utikzyposuop}{-0.8}
      \renewcommand{\Utikzyposport}{-1.6}

      \node[anchor=south] (icap) at (-0.8, \Utikzyposinsn) {\Insns:};
      \node[anchor=south] (ucap) at (-0.8, \Utikzyposuop) {\Uops:};
      \node[anchor=south] (pcap) at (-0.8, \Utikzyposport) {\Ports:};

      \Utikzinsn{1}{0.0}{$i_A$}
      \Utikzinsn{2}{1.5}{$i_B$}

      \Utikzuop{1}{0.0}{$u_A$}
      \Utikzuop{2}{1.5}{$u_B$}

      \Utikzport{1}{0.0}{$p_1$}
      \Utikzport{2}{1.5}{$p_2$}

      \Utikzmapsitou{1}{1}{1}{}
      \Utikzmapsitou{2}{2}{1}{}
      \Utikzmapsutop{1}{1}
      \Utikzmapsutop{2}{2}
    \end{tikzpicture}
    \label{fig:cegpmi_example_first}
  }
  \hspace{1cm}
  \subfloat[]{
    \begin{tikzpicture}
      \renewcommand{\Utikzyposuop}{-0.8}
      \renewcommand{\Utikzyposport}{-1.6}

      \node[anchor=south] (icap) at (-0.8, \Utikzyposinsn) {\Insns:};
      \node[anchor=south] (ucap) at (-0.8, \Utikzyposuop) {\Uops:};
      \node[anchor=south] (pcap) at (-0.8, \Utikzyposport) {\Ports:};

      \Utikzinsn{1}{0.0}{$i_A$}
      \Utikzinsn{2}{1.5}{$i_B$}

      \Utikzuop{1}{0.0}{$u_A$}
      \Utikzuop{2}{1.5}{$u_B$}

      \Utikzport{1}{0.0}{$p_1$}
      \Utikzport{2}{1.5}{$p_2$}

      \Utikzmapsitou{1}{1}{1}{}
      \Utikzmapsitou{2}{2}{1}{}
      \Utikzmapsutop{1}{1}
      \Utikzmapsutop{2}{1}
    \end{tikzpicture}
    \label{fig:cegpmi_example_second}
  }
  \hfill

  \caption{Port mappings that satisfy $\set{([i_A], 1.0), ([i_B], 1.0)}$.}
    \label{fig:cegpmi_example}
  \Description{
    Both port mappings relate two instructions (iA and iB), two micro-ops (uA and uB), and two ports (p1 and p2).
    In both of them, iA is decomposed into one uA micro-op and iB is decomposed into one uB micro-op.
    In the first port mapping, uA can be executed on p1 and uB can be executed on p2.
    In the second port mapping, uA can be executed on p1 and uB can be executed on p1.
  }
\end{figure}
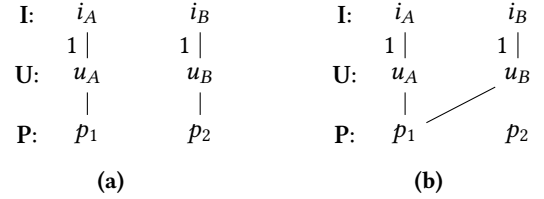

For example, assume an architecture with two instructions $i_A, i_B$, each with a single \uop, and two ports $p_1, p_2$.
So far, we have the following measurements:
\[
  \mathit{Exps} = \set{([i_A], 1.0), ([i_B], 1.0)}
\]
$\mathit{findMapping}(\mathit{Exps})$ might find \cref{fig:cegpmi_example_first} as port mapping $m1$.
However, this is not the only viable port mapping for these measurements:
$\mathit{findOtherMapping}(\mathit{Exps}, m1)$ returns a port mapping~$m2$, \eg, \cref{fig:cegpmi_example_second}.
A distinguishing experiment $\mathit{newExp}$ is $[i_A, i_B]$:
With~$m1$, its inverse throughput is 1.0 cycles per experiment execution while it is 2.0 cycles for~$m2$.

The algorithm is guaranteed to terminate as at least one found port mapping is rejected in each iteration:
No two $\mathit{findOtherMapping}$ calls in a run can yield the same port mapping.
Since there are only finitely many port mappings with only one \uop per instruction, the algorithm only takes a finite number of steps.
In practice, experiments usually rule out more than just a single candidate port mapping.

In the following, we derive SMT-solver-powered implementations of $\mathit{findMapping}$ and $\mathit{findOtherMapping}$ from the linear program in \cref{sec:background:pm_model}.
The idea is to augment the LP such that the port mapping is no longer coded into the constraints, but rather represented in LP variables.
We can then encode $\mathit{findMapping}$ and $\mathit{findOtherMapping}$ as constraints on the LP variables.
An off-the-shelf solver can produce a satisfying model for the variables, from which we decode the result.
Specifically, we formulate a constraint system
\[\mathit{relateThroughput}[\mathit{enc}_{M}, \mathit{enc}_{e}, \mathit{enc}_{t}]\]
parametrized with collections $\mathit{enc}_{M}$,  $\mathit{enc}_{e}$, $\mathit{enc}_{t}$ of variables that represent a port mapping, an experiment, and the experiment's inverse throughput.
We design these constraints such that a model encoding a port mapping~$M$, an experiment~$e$, and a number~$t$ satisfies them if and only if $\tpsim{M}{e} = t$.

\subsubsection{Encoding port mappings and experiments}
\label{ssec:smt:2lvl_encoding}

Experiments are effectively multisets of instruction schemes since their order is irrelevant in the port mapping model.
We represent an experiment as a set $\mathit{enc}_{e} \definedas \set{\mathit{exp}[i] \given i\in \Insns}$ of integer-valued variables.
The value of a variable $\mathit{exp}[i]$ is the number of occurrences of the instruction~$i$ in the experiment.
We constrain these variables to be non-negative.

To represent port mappings, we need variables that encode tripartite graphs between instruction schemes~$\Insns$, ports~$\Ports$, and an unknown set~$\Uops$ of \uops.
Since every considered instruction scheme uses only a single \uop, we chose $\Uops$ such that every instruction~$i$ has its own \uop $u^{(i)}$.\footnote{We can relax this requirement of the algorithm to arbitrary but fixed numbers of \uops by introducing more \uops.}
A set $\mathit{enc}_{M} \definedas \set{\mathit{m}[u^{(i)}, k] \given i\in \Insns, k\in\Ports}$ of boolean variables therefore encodes a port mapping.
When a variable $\mathit{m}[u^{(i)}, k]$ is $\ltrue$ in a model, edges connect instruction~$i$ via \uop $u^{(i)}$ to port~$k$ in the port mapping, \ie, the \uop of $i$~can be executed on~$k$.
\Eg, the port mapping from \cref{fig:cegpmi_example_second} corresponds to a model where $\mathit{m}[u^{(i_A)}, p_1]$ and $\mathit{m}[u^{(i_B)}, p_1]$ are set to $\ltrue$ while $\mathit{m}[u^{(i_A)}, p_2]$ and $\mathit{m}[u^{(i_B)}, p_2]$ are set to $\lfalse$.

We add constraints so that each \uop's number of ports fits the previous throughput measurements.

\subsubsection{Relating port mapping and throughput parametrically}
\label{ssec:smt:formulation}

We adjust the linear program from \cref{sec:background:pm_model} so that the inverse throughput~$t$, the experiment~$e$, and the port mapping $M$ occur only as free variables in the program.

\paragraph{Inverse Throughput}
The inverse throughput is present as a variable~$t$ in the LP, but it is not free:
Constraints A-E of the LP only assert that the experiment can be executed according to the port mapping within at most~$t$ cycles and the minimization objective ensures that~$t$ corresponds to an optimal \uop distribution.
If the value of~$t$ was fixed by a constraint, the minimization objective would effectively be disabled.
We therefore replace the optimization objective of the LP with more constraints that ensure optimality of the execution.
We use SMT formulas in the theory of linear integer and real arithmetic (LIRA) to obtain a more intuitive formulation with logical disjunctions and implications:%
  \footnote{The multiplications in constraint~\ref{bnI} all involve either a boolean variable or a constant and are therefore encodable without undecidable theories.}
\begin{alignat*}{2}
  &  \displaystyle\biglor_{k\in\Ports} q_k          &&\tag{F}\label{bnF}\\
  &  q_k \leftrightarrow (p_k = t)         && \fspace\text{for all ports $k\in\Ports$}\tag{G}\label{bnG}\\
  &  j_u \rightarrow q_k                   && \fspace\text{if $(u,k)\in E$}\tag{H}\label{bnH}\\
  &  \displaystyle\sum_{u\in\Uops} j_u \cdot \hspace{-0.8em}\displaystyle\sum_{(i, n, u)\in F} \hspace{-1.0em}\insnf{e}(i) \cdot n = \displaystyle\sum_{k\in\Ports} q_k \cdot t &\tag{I}\label{bnI}
\end{alignat*}

These constraints are based on a result by Ritter and Hack~\citep[Section 4.5]{ritter2020pmevo}:
A distribution to ports is optimal if and only if there is a non-empty set~$Q$ of bottleneck ports that are all utilized for the full number of cycles with \uops that can only be executed on ports in~$Q$.
In the formulas, a port~$k$ is in the set~$Q$ of bottleneck ports iff the boolean variable~$q_k$ is \ltrue.
Constraint~\ref{bnF} asserts that~$Q$ is not empty.
With constraint~\ref{bnG}, we ensure that each bottleneck port is utilized for the full $t$~cycles.
The boolean $j_u$~variables encode a set~$J$ of \uops~$u$ that can only be executed on bottleneck ports in~$Q$, as enforced by constraint~\ref{bnH}.
Lastly, constraint~\ref{bnI} ensures that only \uops from~$J$ contribute to the utilization of the ports in~$Q$.

\paragraph{Experiment and Port Mapping}
To introduce the experiment encoding, we use the $\mathit{exp}[i]$~variables instead of the fixed numbers~$\insnf{e}(i)$ of occurrences for each instruction~$i$.
We replace the constraints~\ref{js3E} and~\ref{bnI} to integrate the port mapping encoding into the constraints with logical implications:
\begin{align*}
  \begin{multlined}
    x_{uk} = 0 \quad\text{if $ (u, k) \not\in E$} \hfill (E)\\
    \leadsto \fspace \lnot \mathit{m}[u, k] \rightarrow x_{uk} = 0 \quad\text{for all $u\in\Uops$, $k\in\Ports$}
  \end{multlined}\\
  \begin{multlined}
     j_u \rightarrow q_k \quad\text{if $(u,k)\in E$}  \hfill (I)\\
      \leadsto \fspace \mathit{m}[u, k] \rightarrow ~(j_u \rightarrow q_k) \quad\text{for all $u\in\Uops$, $k\in\Ports$}
  \end{multlined}
\end{align*}
The resulting constraints form $\mathit{relateThroughput}$, with inverse throughput, experiment, and port mapping as free variables.

\subsubsection{$\mathit{findMapping}$ and $\mathit{findOtherMapping}$}
\label{ssec:smt:2lvl_algo_components}

The parametric $\mathit{relateThroughput}[\mathit{enc}_{M}, \mathit{enc}_{e}, \mathit{enc}_{t}]$ constraints make the functions from \cref{algo:cegpminf} straightforward to implement:

$\mathit{findMapping}(\mathit{Exps})$ uses a single free port mapping encoding $M_{\mathit{free}}$.
For each experiment $e$ with inverse throughput $t_e$, we assert $\mathit{relateThroughput}$ constraints for $M_{\mathit{free}}$ and fresh experiment and throughput encodings that are hardwired to $e$ and $t_e$, respectively:
\begin{align*}
  \varphi_{\mathit{findMapping}} \definedas &~\bigland_{(e, t_e)\in\mathit{Exps}} \mathit{relateThroughput}[M_{\mathit{free}}, e, t_e]
\end{align*}
The resulting conjunction ensures that the port mapping encoded in a satisfying model yields the observed inverse throughput for all experiments.
We use an off-the-shelf SMT solver to check for satisfiability.
If the constraints are unsatisfiable, the observed throughputs cannot be explained in the model and we find no mapping.
Otherwise, we extract and return the port mapping from the values of the encoding variables in the satisfying model produced by the solver.

$\mathit{findOtherMapping}(M_1, \mathit{Exps})$ also uses these constraints to require that the found port mapping satisfies the experiments, and adds more:
Another mapping encoding is hard-wired to the input port mapping~$M_1$.
For a free experiment encoding $e_{\mathit{free}}$, we use two more instances of the \textit{relateThroughput} constraints to encode the inverse throughputs of both port mapping encodings in two variables $t_1, t_2$.
Lastly, we assert that $t_1 \neq t_2$, \ie, the experiment distinguishes the hard-wired and the free port mapping:
\begin{align*}
  \varphi_{\mathit{findOther}} \definedas &~t_1 \neq t_2 \land \mathit{relateThroughput}[M_1, e_{\mathit{free}}, t_1]\\
    &\hspace*{-1cm}\land~\mathit{relateThroughput}[M_{\mathit{free}}, e_{\mathit{free}}, t_2] \land \varphi_{\mathit{findMapping}}
\end{align*}

\subsubsection{Addressing Benchmarking Limitations}
\label{ssec:smt:practical_extensions}

When benchmarking modern processors, inexact measurements due to noise and errors are inevitable.
We therefore adapt the constraints:
A parameter~$\varepsilon$ constrains the maximal difference between measured and modeled cycles per instruction (CPI) of the experiments.\footnote{\ie, inverse throughput divided by the length of the experiment.}
The following constraint encodes equality of the measured and modeled inverse throughputs $t_e$ and $\mathit{enc}_{t}$:
\[
\abs{\mathit{enc}_{t} - t_e} < \varepsilon \cdot \abs{\mathit{exp}}
\]

When asserting that the modeled inverse throughputs of the two port mappings in $\mathit{findOtherMapping}$ are different, no observed value may be considered equal to both modeled inverse throughputs.
Otherwise a found experiment might not rule out any of the candidate mappings.
This can be guaranteed if the modeled CPIs differ by at least $2\cdot \varepsilon$:
\begin{gather*}
  \abs{t_1 - t_2} > 2\cdot\varepsilon \cdot\abs{\mathit{exp}}
\end{gather*}

Moreover, \textit{findOtherMapping} can produce excessively large microbenchmarks that hit bottlenecks outside the port mapping, \eg, the cache boundaries.
We therefore follow a stratified approach:
First, we only search distinguishing experiments with a single instruction and increase this bound once no more experiments are found.
If increasing the bound yields no experiments, we run \textit{findOtherMapping} without bound.
If this produces no experiment either, the algorithm terminates; otherwise it continues with a larger bound.
As a result, the benchmark experiments have minimal size without sacrificing the completeness of the algorithm.

\subsection{Handling Pipeline Bottlenecks}
\label{relaxed:ssec:bottlenecks}

The port mapping model assumes that throughput is only limited by the availability of functional units.
In practice, that is not the case.
All modern processors that we are aware of, including recent designs by Intel and AMD, cannot sustain a full utilization of all ports.
The culprit is often the decoding frontend (including the caches) or the instruction retirement rate.
When a bottleneck limits the execution to at most~$\bnsym$ instructions per cycle, experiments that are faster according to the port mapping model are slowed to meet the limit.

Such bottlenecks can affect the correctness of our algorithm.
The checks for equivalence of blocking instructions (\cref{relaxed_algo:ssec:find_blocking_insns}) and for evading \uops (\cref{relaxed_algo:ssec:evading_insns}) measure if an experiment utilizes more than a certain number $n$ of ports.
These checks fail if there is no gap between $\bnsym$ and the largest such~$n$, \ie, the maximal port set size of any \uop.
We need to check this requirement when applying the algorithm.

The counter-example-guided inference algorithm in \cref{relaxed_algo:ssec:cegpmi} needs an adjustment for such bottlenecks:
We change $\mathit{relateThroughput}[\mathit{enc}_{M}, \mathit{enc}_{e}, \mathit{enc}_{t}]$ such that $\mathit{enc}_{t}$ is the maximum of the number $\tpsim{M}{e}$ of cycles according to the model and the peak inverse throughput $\abs{e}/\bnsym$ at the bottleneck.
Some theoretically distinguishable port mappings become indistinguishable with this adjustment.

\subsection{Supported Microarchitectures}
\label{ssec:requirements}

Our algorithm has the following requirements:
\begin{itemize}[leftmargin=*]
  \item We need to measure the number of cycles required to execute a piece of code.
    Such functionality is commonplace in contemporary Intel, AMD, and ARM microarchitectures.
  \item There needs to be a counter for the total number of \uops executed for a piece of code.
    Recent Intel Core architectures support this, and AMD's Zen, Zen+, and Zen2 are documented to support this as well.\footnote{See \cref{relaxed:ssec:eval:singletons} for more on this and subsequent Zen microarchitectures.}
  \item The processor's throughput bottleneck should not be hit when executing only instructions of the same kind.
    AMD's Zen-family microarchitectures (up to Zen4) satisfy this requirement \citep{agneruarch}, most Intel Core architectures violate it.
\end{itemize}
According to available documentation, examples of contemporary microarchitectures as of 2024 that satisfy these requirements include:
\begin{itemize}[leftmargin=*]
  \item AMD's Zen, Zen+, and Zen2 microarchitectures:
    They sustain a throughput of 5 instructions per cycle (IPC) \citep[Chapter 22]{agneruarch} and individual \uops have up to 4 ports available.
    The \enquote{Retired Uops} counter~\citep[Section 2.1.15.4.5]{amd-zenp-ppr} is documented to count \uops.\footnote{Our experiments indicate that this performance counter behaves like the \enquote{Retired Ops} counter of Zen3 and Zen4~\citep[Section 2.1.15.4.5]{amd-zen3-ppr}.
      With a throughput of 6 IPC and up to 4 ports per \uop~\citep[Chapters 23--24]{agneruarch}, Zen3 and Zen4 could therefore be handled similarly as the previous Zen generations.}
  \item Intel's Golden Cove microarchitecture, which is used in their Sapphire Rapids and Alder Lake processors:
    It sustains 6 IPC \citep[Section 13.1]{agneruarch} and its \uops have up to 5 ports available~\citep{abel19uopsinfo}.
    Golden Cove has a \texttt{UOPS\_EXECUTED.THREAD} performance counter~\citep{intel-perfmon}.
  \item Fujitsu's A64FX microarchitecture:
    Its decoder can issue 4 instructions per cycle and \uops can use up to 3 ports~\citep[Chapter 2]{fujitsu-man}.
    This architecture provides a \texttt{UOP\_SPEC} performance counter~\citep[p.\,16]{fujitsu-pmu}.
  \item ARM's Neoverse V2:
    Its decoder sustains a throughput of 8 IPC while \uops can use up to 6 ports~\citep[Section 2.1]{neoverse-optguide}.
    Its \texttt{OP\_RETIRED} counter~\citep[Table 18-1]{neoverse-refman} counts executed \uops.
  \item Apple's M1:
    According to results by Dougall Johnson~\citep{johnson21m1}, its performance core sustains 8 IPC and each \uop has up to 6 ports available.
    Johnson uses an an undocumented performance counter to count \uops.
\end{itemize}

Additionally, as in the original \uopsinfo algorithm, we assume that there is a blocking instruction for most \uops of the microarchitecture.
Where this requirement is not met, replacement instructions with a throughput dominated by the respective \uop need to be specified manually.
In the following case study, we show that, contrary to official documentation, such \uops occur in AMD's Zen+ microarchitecture.

\section{Case Study: The AMD Zen+ Architecture}
\label{relaxed:sec:eval}

\newcommand{\numSchemesInitial}{2,980}
\newcommand{\numSchemesAfterCharacterization}{2,323}
\newcommand{\numBlockingInsnCandidatesAfterCharacterization}{691}
\newcommand{\numSchemesAfterEquivCheck}{1,887}
\newcommand{\numSchemesAfterSMT}{1,819}
\newcommand{\numBlockingInsnCandidates}{563}
\newcommand{\numBlockingInsnCandidatesWithoutUopsInfoData}{266}
\newcommand{\numUniqueBlockingInsns}{13}
\newcommand{\numSchemesWithMapping}{1,700}
\newcommand{\numSchemesInEval}{577}
\newcommand{\numSchemesWithMappingWithoutUopsinfoData}{1,142 (67\%)}

\noindent
We evaluate our port mapping inference algorithm with the AMD Zen+ microarchitecture.
This allows us to use \pmevo~\citep{ritter2020pmevo} and \palmed~\citep{derumigny22palmed} as points of comparison in \cref{ssec:eval:accuracy}.
Since Zen+ does not have full per-port \uop counters, the original \uopsinfo algorithm is not applicable.
We also compare our results to the available documentation for Zen+:
\begin{itemize}[leftmargin=*]
  \item AMD's Software Optimization Guide (SOG) \citep{amd-zenp-opt-manual} describes the microarchitecture and documents instruction latencies and throughputs, if they are microcoded, and, for simple instructions, their execution units.
  \item Agner Fog's microarchitecture guide \citep{agneruarch} analyzes the similar Zen architecture based on manual microbenchmarks.
  \item Fog \citep{agner} and \uopsinfo \citep{abel19uopsinfo} provide tables with measured latencies, throughputs, and numbers of \uops of individual instructions.
    They include the port usage of floating point (FP)/vector instructions since per-port performance counters for these units are available \citep[Section 2.1.15.4.1]{amd-zenp-ppr}.
  \item WikiChip collects information on Zen and Zen+, in part from AMD's marketing resources~\citep{wikichipZen,wikichipZenp}.
\end{itemize}

Our test system has an AMD Ryzen 5 2600X processor and \unitgigabyte{32} of RAM.
It runs the port mapping inference algorithm and automatically performs microbenchmarks when required.
Simultaneous multi-threading and frequency scaling are disabled.
We measure inverse throughput and retired \uops with a technique similar to \nanobench \citep{abel20nanobench}, taking the median over 11 repeated microbenchmark runs.\footnote{We use the \texttt{PMCx0C1} (\enquote{Retired Uops}) counter \citep[Section 2.1.15.4.5]{amd-zenp-ppr}.}
We consider two throughput measurements equal if the implied cycles per instruction (CPI) differ by at most $\varepsilon = \relaxedalgoeps$.
This value allows us to distinguish if experiments use five ports ($0.20$ CPI) or four ports ($0.25$ CPI).
Zen+ meets our bottleneck requirement: Five blocking instructions can be executed per cycle \citep[Section 22.21]{agneruarch} and \uops have at most four ports.

We take the x86-64 instruction schemes from \uopsinfo and remove control flow and system instructions as well as instructions with known input-dependent performance characteristics.
For FP and vector operations, we only consider instructions from the AVX and AVX2 instruction set extensions.
This gives us \numSchemesInitial{} instruction schemes.
We further reduce this set of instruction schemes when the need arises throughout the stages of the algorithm.

\subsection{Identifying Blocking Instruction Candidates}
\label{relaxed:ssec:eval:singletons}

The first algorithm stage benchmarks every single instruction under investigation individually.
Instructions that are executed with a single \uop are blocking instructions.

\subsubsection{Counting \uops}
Compared to AMD's documentation, we measure unexpected numbers of \uops, \eg, for this instruction scheme:
\begin{center}
  \ischeme{add\opgpr{32}{RW},\wordptrprefix{d}\opmem{32}{R}}
\end{center}
It loads a value from memory, adds it to the value of a register, and writes the result into the same register.
According to AMD's SOG \citep[Table 1]{amd-zenp-opt-manual}, this instruction uses two \uops: one that loads and one that adds.
However, the \enquote{Retired Uops} counter only increases by one.
We observe the same for any instruction with memory operands.
\uopsinfo and Fog's tables, which also rely on this performance counter, agree with our observations.
While our inquiry with AMD's support remains unanswered, there is evidence that the \enquote{Retired Uops} performance counter, \texttt{PMCx0C1}, counts \emph{macro-ops} instead of \uops:
The observed values are consistent with the SOG's macro-op numbers and AMD's documentation for the more recent Zen 3 and 4 microarchitectures \citep[Section 2.1.15.4.5]{amd-zen3-ppr} documents this identifier for counting macro-ops.

AMD's macro-ops are a representation between x86-64 instructions and the \uops that are executed by the execution units \citep[Section 2.3]{amd-zenp-opt-manual}.
Many instructions are implemented with a single macro-op, whereas, \eg, 256-bit-wide vector operations use two narrower macro-ops.
Complex instructions are microcoded with a greater number of macro-ops.

To the best of our knowledge, there is no detailed published information on how macro-ops are decomposed into \uops and no suitable performance counter for experimental characterization.
Since our algorithm requires a count of \uops, we postulate a macro-op-to-\uop correspondence in the Zen+ microarchitecture, based on AMD's SOG \citep[Section 2.3]{amd-zenp-opt-manual}:
Let~$n$ be the number of macro-ops observed when executing a basic block $\mathit{bb}$.
We obtain the \uop count by adding
\begin{itemize}[leftmargin=*]
  \item 1 for each memory operand with a width of at most 128~bits (excluding \enquote{load effective address} and loading \ischeme{mov}s),
  \item 2 for each memory operand with a width of 256~bits (as they are implemented as two 128-bit operations).
\end{itemize}

For one case, we deviate from the SOG:
It claims that \ischeme{mov}s which \emph{store} to memory do not require an additional \uop.
This contradicts our observations:
\begin{itemize}[leftmargin=*]
  \item
    A store-\ischeme{mov} together with four simple register-additions takes 1.25 cycles.
    Therefore, it has a \uop that is restricted to the four ALU ports.
  \item
    A \ischeme{vmovapd} vector-register-to-memory store (documented with a store \uop and one to deliver the stored data) together with the four additions takes only 1.0 cycles.
    Hence, no \uops of this instruction are restricted to the ALU ports.
  \item
    A storing \ischeme{mov} with a storing \ischeme{vmovapd} leads to an inverse throughput of 2 cycles.
    These instructions therefore interfere, \ie, a \uop of the \ischeme{mov} instruction uses a port that the \ischeme{vmovapd} instruction also needs.
\end{itemize}
Hence, the storing \ischeme{mov} instruction has a \uop that is restricted to one or more ALU ports and one for a non-ALU port.
Therefore, similar to Intel architectures \citep[Section 5.1.1]{abel19uopsinfo}, there is no proper blocking instruction for memory store \uops.

\subsubsection{Problematic Instructions.}

For several instruction schemes, we observe breaks in the algorithm's assumptions:
\begin{itemize}[leftmargin=*]
  \item \ischeme{nop}s and 32 or 64-bit-wide register-to-register \ischeme{mov}s use no ports:
    The processor resolves such \ischeme{mov}s via register renaming~\citep[Section 22.13]{agneruarch} and implements \ischeme{nop}s without \uops.
    No port mapping is necessary for these cases.

  \item Some FP instructions execute slower than the port mapping model permits, \eg, divisions, square-root computations and approximate reciprocals.

  \item A \ischeme{mov} of a 64-bit immediate into a GPR causes unreliable measurements.
    As these constants are unusual in the ISA, they use special handling in the hardware \citep[Section 2.9]{amd-zenp-opt-manual}.

  \item We cannot measure instructions that modify operands that are hardwired or restricted to \texttt{ah}--\texttt{dh} registers without observing effects of data dependencies.

\end{itemize}
We exclude all these instruction schemes, leaving \numSchemesAfterCharacterization{} remaining schemes.
Of these, \numBlockingInsnCandidatesAfterCharacterization{} are identified as blocking instruction candidates.

\subsection{Filtering Equivalent Blocking Instructions}

Next, we run microbenchmarks for pairs of blocking instruction candidates with equally-sized port sets to check if they are equivalent.
We encounter further problematic instructions in these experiments:
Conditional move instructions, AES de/encryption operations, numerical conversions of the \texttt{vcvt*} family, and double-precision FP multiplications cause unstable measurements when benchmarked with other instructions.
FP/vector operations with three read operands, like fused multiply-and-add instructions and some vector blending operations, do not fit the port mapping model either in Zen+.
While these operations can execute on two of the four ports of the FP unit, they use data lines of a third port~\citep[Section 2.11]{amd-zenp-opt-manual}.
This third port meanwhile has to idle, which we observe as contradicting equivalence information.
We exclude these instructions from the following steps.

This leaves us with \numSchemesAfterEquivCheck{} instruction schemes in total, with \numBlockingInsnCandidates{} blocking instruction candidates.
Of these candidates, \numUniqueBlockingInsns{} are identified as unique blocking instructions.
\cref{tab:relaxed_algo:blocking_insns} shows them with the number of candidates per equivalence class.

This selection is consistent with \uopsinfo:
If we found two candidates equivalent and if they are covered by \uopsinfo, then their reported port usages are equal.
\uopsinfo does not cover \numBlockingInsnCandidatesWithoutUopsInfoData{} of our \numBlockingInsnCandidates{} candidates.
AMD's tables do not agree for 33 instruction schemes.
\Eg, they document the following \texttt{xmm} vector comparisons with the same two ports:
\begin{center}
  \ischeme{vpcmp\textbf{gtq}} \quad \ischeme{vpcmp\textbf{eqq}} \quad \ischeme{vpcmp\textbf{gtb}}
\end{center}
In our measurements, only the second (testing equality for $2\times64$-bit integers) has two ports, whereas the first and third have one and three ports available (greater-than tests for $2\times64$-bit and $16\times8$-bit integers, respectively).
Fog's table and \uopsinfo agree with our measurements; this appears to be an error in AMD's documentation.

\begin{table}
  \small
  \caption{Identified blocking instruction classes for AMD Zen+. Representants were selected manually for clarity.}
  \label{tab:relaxed_algo:blocking_insns}
  \centering
  \begin{tabular}{rcl}
    \# Ports \hfill Instruction Scheme & \hspace*{-2em}\# Equiv.\hspace*{-2em} & Description\\
    \toprule
    4 \hfill \ischeme{add\opgpr{32}{RW},\opgpr{32}{R}} &  242 & ALU ops\\
    \ischeme{vpor\opxmm{W},\opxmm{R},\opxmm{R}} &  21 & logical vector ops\\ 
    \midrule
    3 \hfill \ischeme{vpaddd\opxmm{W},\opxmm{R},\opxmm{R}} &  30 & vector int. arith.\\ 
    \midrule
    2 \hfill \ischeme{vminps\opxmm{W},\opxmm{R},\opxmm{R}} &  143 & FP compare, mul. \\ 
    \ischeme{vbroadcastss\opxmm{W},\opxmm{R}} &  50 & vector layouting\\ 
    \ischeme{vpaddsw\opxmm{W},\opxmm{R},\opxmm{R}} &  17 & saturating vec.\ ops\\ 
    \ischeme{vaddps\opxmm{W},\opxmm{R},\opxmm{R}} &  10 & FP additions\\ 
    \ischeme{mov\opgpr{32}{W},\wordptrprefix{d}\opmem{32}{R}} &  6 & memory loads\\
    \midrule
    1 \hfill \ischeme{vpslld\opxmm{W},\opxmm{R},\opxmm{R}} &  27 & vector shifts\\ 
    \ischeme{vpmuldq\opxmm{W},\opxmm{R},\opxmm{R}} &  10 & elaborate vec.\ mul.\\ 
    \ischeme{imul\opgpr{32}{RW},\opgpr{32}{R}} &  9 & integer mul.\\
    \ischeme{vroundps\opxmm{W},\opxmm{R},\opimm{8}} &  4 & vector rounding\\ 
    \ischeme{vmovd\opxmm{W},\opgpr{32}{R}} &  2 & vector-to-GPR mov\\
  \end{tabular}
\end{table}

\subsection{Computing a Mapping for the Blocking Instructions}
\label{relaxed_uops:ssec:case_study_smt}
Here, we compute a port mapping for the blocking instructions with the counter-example-guided inference algorithm.
We use \texttt{z3}~\citep{z3} (V.4.12.1) as SMT solver and select $\varepsilon = \relaxedalgoeps$.
Following the SOG \citep{amd-zenp-opt-manual}, we use a set of 10 ports.

As there are no proper blocking instructions for store operations, we add \enquote{improper} blocking instructions manually:
\begin{itemize}[leftmargin=*]
  \item \ischeme{mov\wordptrprefix{d}\opmem{32}{W},\opgpr{32}{R}}, which stores a 32-bit value from a general purpose register into memory, and
  \item \ischeme{vmovapd\wordptrprefix{xmm}\opmem{128}{W},\opxmm{R}}, which stores a 128-bit value from a vector register into memory.
\end{itemize}
While we expect to use only the \ischeme{mov} instruction in place of a blocking instruction for the store \uop, both are required to infer that the store \uop does not use an ALU instruction, \cf \cref{relaxed:ssec:eval:singletons}.
We augment the SMT formulas such that the improper blocking instructions use exactly two \uops and one of their \uops is equal to one with a proper blocking instruction.
These constraints avoid prohibitively long execution times.

For three blocking instructions, the generated experiments exhibit throughputs outside of the port mapping model:
\begin{itemize}[leftmargin=*]
  \item The \ischeme{imul} scheme for scalar integer multiplications, \eg, when combined with four additions:
    \begin{center}
      $4 \times$ \ischeme{add\opgpr{32}{RW},\opgpr{32}{R}}\\
      $1 \times$ \ischeme{imul\opgpr{32}{RW},\opgpr{32}{R}}
    \end{center}
    Since \texttt{add} has four ports and \texttt{imul} is restricted to one, two inverse throughputs are possible in the port mapping model:
    1.25 cycles, if \texttt{imul} uses a port of the \texttt{add} instruction or 1.0 cycles, if their ports are disjoint.
    While AMD's SOG \citep[Section 2.10.2]{amd-zenp-opt-manual} indicates the former, we measure ca.\ 1.5 cycles for this experiment, matching neither case.

  \item \ischeme{vpmuldq}, which represents uncommon vector multiplication operations,\footnote{This specific instruction multiplies the 32-bit integers at even-numbered lanes in the source registers without overflows into a vector of 64-bit integers.} leads to experiments that run slower than their port usage would imply.
    This deviation from the modeled throughputs would require a larger $\varepsilon$, reducing the accuracy for other instructions.

  \item For \ischeme{vmovd}, we observe inconsistent resource conflicts when combined with different instructions.
    As this instruction scheme is untypical in that it transfers data between vector registers and the GPRs, its throughput might depend on resources outside of the port mapping model.
\end{itemize}
These instructions cause $\mathsf{UNSAT}$ results in the $\mathit{findMapping}$ method.
We exclude them and instructions with the same mnemonics (as we expect them to share aspects of the problematic instructions) from this investigation.

\begin{table}
  \caption{Documented and inferred port usage of the blocking instructions for Zen+. Inferred ports were renamed to ease comparison. }
  \label{tab:relaxed_algo:blocking_insn_mapping}
  \centering
  \small
  \begin{tabular}{cll}
    \toprule
    Instruction Scheme & Doc.\ Ports\hspace*{-1ex} & Inferred Ports\\
    \toprule
    \ischeme{add\opgpr{32}{RW},\opgpr{32}{R}} & ALU & [6,7,8,9] \\
    \ischeme{vpor\opxmm{W},\opxmm{R},\opxmm{R}} & FP 0,1,2,3\hspace*{-2ex} & [0,1,2,3] \\ 
    \midrule
    \ischeme{vpaddd\opxmm{W},\opxmm{R},\opxmm{R}} & FP 0,1,3 & [0,1,3] \\ 
    \midrule
    \ischeme{vminps\opxmm{W},\opxmm{R},\opxmm{R}} & FP 0,1 & [0,1] \\ 
    \ischeme{vbroadcastss\opxmm{W},\opxmm{R}} & FP 1,2 & [1,2] \\ 
    \ischeme{vpaddsw\opxmm{W},\opxmm{R},\opxmm{R}} & FP 0,3 & [0,3] \\ 
    \ischeme{vaddps\opxmm{W},\opxmm{R},\opxmm{R}} & FP 2,3 & [2,3] \\ 
    \ischeme{mov\opgpr{32}{W},\wordptrprefix{d}\opmem{32}{R}} & AGU & [4,5] \\
    \midrule
    \ischeme{vpslld\opxmm{W},\opxmm{R},\opxmm{R}} & FP 2 & [2] \\ 
    \ischeme{vroundps\opxmm{W},\opxmm{R},\opimm{8}}\hspace*{-1ex} & FP 3 & [3] \\ 
    \midrule
    \ischeme{mov\wordptrprefix{d}\opmem{32}{W},\opgpr{32}{R}} & AGU & [5] + [6,7,8,9]\\
    \ischeme{vmovapd\wordptrprefix{xmm}\opmem{128}{W},\opxmm{R}} & FP 2 & [5] + [2]\\
    \bottomrule
  \end{tabular}
\end{table}

In three runs with the remaining blocking instructions, the algorithm terminated within 12--20 hours after generating 55--59 experiments with up to five instructions.
\cref{tab:relaxed_algo:blocking_insn_mapping} shows the inferred port mapping and the documented port usage.
For vector and FP instructions, where documented port usages are available, our port mapping is equivalent.

Results for the \ischeme{add} blocking instruction differ across repeated algorithm runs in whether a port is shared with the FP instructions:
Besides the mapping in \cref{tab:relaxed_algo:blocking_insn_mapping}, \enquote{[6,7,8,9]}, variants like \enquote{[0,6,7,8]} and \enquote{[1,6,7,8]} that use FP ports are possible.
These variants are indistinguishable with the throughput bottleneck of five instructions per cycle.
Which result we get depends on choices of the SMT solver.
This ambiguity would be resolved with a less tight bottleneck or with blocking instructions for the individual FP ports or fine-grained subsets of the ALU ports.
We use \enquote{[6,7,8,9]} in the rest of the algorithm as it is consistent with the documentation.

The results for the improper blocking instructions (at the bottom of the table) are consistent with the expectations:
They have a \uop (presumably for storing to memory) for port 5 in common.
\ischeme{vmovapd} has an additional \uop for port~2, which \uopsinfo reports as its port usage.
For \ischeme{mov}, the additional \uop is an ALU \uop, matching our observations from \cref{relaxed:ssec:eval:singletons}.

\subsection{Computing the Remaining Port Mapping}
\label{ssec:relaxed_algo:eval:remaining_mapping}

\begin{figure*}
  \subfloat[Accuracy Metrics]{
    \label{tab:relaxed_algo:pred_eval}
    \small
    \centering
    \raisebox{2.6cm}{
    \begin{tabular}{cccc}
      \toprule
      & MAPE & PCC & $\tau_K$ \\
      \midrule
      \pmevo & 28.0\% & 0.83 & 0.72 \\
      \palmed & 35.2\% & 0.79 & 0.66 \\
      Ours & 6.6\% & 0.96 & 0.90 \\
      \bottomrule
    \end{tabular}
    }
  }
  \subfloat[PMEvo]{
    \label{fig:heatmap:pmevo}
    \hspace*{0.0cm}\includegraphics[width=0.24\textwidth]{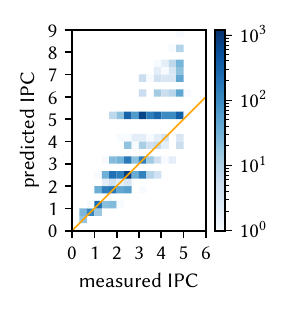}
  }
  \hspace*{-0.2cm}
  \subfloat[Palmed]{
    \label{fig:heatmap:palmed}
    \includegraphics[width=0.24\textwidth]{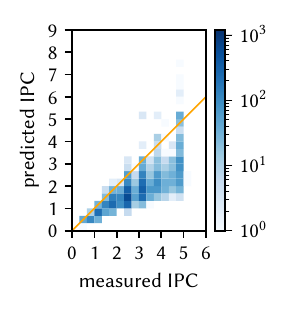}
  }
  \hspace*{-0.2cm}
  \subfloat[ours]{
    \label{fig:heatmap:ours}
    \includegraphics[width=0.24\textwidth]{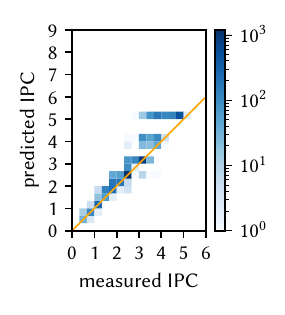}
  }
  \caption{IPC prediction accuracy for Zen+ in metrics (a) and as heatmaps of predicted vs.\ measured IPC per model (b-d).}
  \label{fig:heatmap}
  \Description{
    Subfigure (a) shows a table with the mean absolute percentage error (MAPE), Pearson's correlation coefficient (PCC), and Kendall's tau for the three models.
    PMEvo has a MAPE of 28.0\%, a PCC of 0.83, and a Kendall's tau of 0.72.
    Palmed has a MAPE of 35.2\%, a PCC of 0.79, and a Kendall's tau of 0.66.
    Our model has a MAPE of 6.6\%, a PCC of 0.96, and a Kendall's tau of 0.90.
    Subfigures (b-d) show heatmaps of predicted vs. measured IPC per model.
    The maximal measured IPC is 5.
    For PMEvo, both over- and underestimations of the measured IPC are common.
    For experiments with a measured IPC of 2 or more, the IPC predictions are often too high, with many predictions of circa 5 IPC.
    Few predictions exceed 5 IPC, ranging up to 9 IPC.
    For Palmed, the IPC estimates are often too low.
    The estimates usually range between the measured IPC and a third of the measured IPC. For instance, experiments with a measured IPC of 5 are predicted between 1.5 and 5 IPC.
    Almost no overestimations are visible.
    For our model, the predictions are substantially closer to the measured IPC.
    Slight overestimations of the IPC are more common than underestimations.
    For experiments with a measured IPC of at most 3, allmost all predictions are less than 1 IPC off.
    For experiments with a higher measured IPC, the estimates of up to 5 IPC occur.
  }
\end{figure*}

\noindent
Finally, the algorithm benchmarks the remaining instructions against the suite of blocking instructions.
To combat unstable measurements, we run this part of the algorithm three times and only report the port usage for an instruction if at least two of the runs agree.
We use \ischeme{mov\wordptrprefix{d}\opmem{32}{W},\opgpr{32}{R}} to block the store port 5.

The results follow regular patterns for most instructions:
\begin{itemize}[leftmargin=*]
  \item 256-bit wide AVX instructions use \uops of the same kinds as the 128-bit variants, only with twice the number, \eg:

    \begin{center}
      \begin{tabular}{rl}
        \small\ischeme{vpcmpeqq\opxmm{W},\opxmm{R},\opxmm{R}} \hspace*{-1em} &$\leadsto  1\times[0, 3]$ \\
        \small\ischeme{vpcmpeqq\opymm{W},\opymm{R},\opymm{R}} \hspace*{-1em} &$\leadsto  2\times[0, 3]$ \\
      \end{tabular}
    \end{center}

  \item Instruction schemes with a read memory operand differ from their register-only counterparts by one load \uop (two for double-pumped 256-bit AVX instructions), \eg:
    \begin{center}
      \begin{tabular}{rl}
        \small\ischeme{add\opgpr{32}{RW},\opgpr{32}{R}} \hspace*{-1em} & $\leadsto  [6, 7, 8, 9]$ \\
        \small\ischeme{add\opgpr{32}{RW},\wordptrprefix{d}\opmem{32}{R}} \hspace*{-1em} & $\leadsto [6, 7, 8, 9] + [4,5]$ \\
      \end{tabular}
    \end{center}

    This follows our postulated macro-op decomposition.

  \item Simple scalar instructions with a read and written memory operand use an ALU \uop and a store \uop:
    \begin{center}
      \begin{tabular}{rl}
        \small\ischeme{add\opgpr{32}{RW},\opgpr{32}{R}} \hspace*{-1em} & $\leadsto  [6, 7, 8, 9]$ \\
        \small\ischeme{add\wordptrprefix{d}\opmem{32}{RW},\opgpr{32}{R}} \hspace*{-1em} & $\leadsto  [6, 7, 8, 9] + [5]$ \\
      \end{tabular}
    \end{center}
    \noindent
    In contrast to Intel architectures, Zen+ has no separate \uops for the two memory operations in read-modify-write instructions.
    As an exception, operations on $\leq 32$ bit use an additional \uop on the address generation units [4,5].
\end{itemize}
Overall, 70\% of the remaining \numSchemesAfterSMT{} considered instruction schemes fall into this category.

For complex instructions, we find unexpected results, \eg:
\begin{align*}
  &\begin{multlined}
    \ischeme{bsf\opgpr{64}{W},\wordptrprefix{q}\opmem{64}{R}} \\ \leadsto  9\times[6, 7, 8, 9] + [4, 5] + 9\times [0, 1, 2, 3]
  \end{multlined}\\
  &\begin{multlined}
    \ischeme{vphaddw\opxmm{W},\opxmm{R},\opxmm{R}}\\ \leadsto  [0, 1, 2, 3] + [0, 1, 3] + 2\times [1, 2] + 4\times[6,7,8,9]
  \end{multlined}
\end{align*}
The former is a \emph{bit scan forward} instruction, which finds the position of the least significant bit set in its read (memory) operand.
The latter is a horizontal vector addition.
Their inferred port usages are unexpected in two ways:
They contain more \uops than reported by the performance counter (8+1 counted and adjusted for a memory operand for \ischeme{bsf} and 4 counted for \ischeme{vphaddw}) and they include \uops for unlikely ports.
For the scalar integer operation \ischeme{bsf}, we do not expect vector/FP ports $[0, 1, 2, 3]$, whereas the vector operation \ischeme{vphaddw} is unlikely to use the scalar ALU ports $[6, 7, 8, 9]$.
We suspect these to be spurious observations caused by the processor's microcode sequencer (MS).
For instructions with many \uops, the processor's instruction decoder only emits an entry point address for the MS ROM.
The MS then emits the relevant operations.
Our observations match a MS that emits four operations per cycle while stalling the remaining frontend.
Rather than \uops that cannot execute on unblocked ports, we measure the overhead of this bottleneck.
This occurs for 8\% of the \numSchemesAfterSMT{} considered instruction schemes.

For 7\% of the instruction schemes, \eg, for bit shift operations on vector registers, the experiments yield throughputs that are unstable or outside the port mapping model.

This last stage of the algorithm takes 8--10 hours.
The last two stages of the algorithm dominate the running time of the algorithm, with a total of 20--28 hours.
Overall, we inferred a port mapping for \numSchemesWithMapping{} of the initial \numSchemesInitial{} instruction schemes.
\uopsinfo has no port mapping for \numSchemesWithMappingWithoutUopsinfoData{} of these \numSchemesWithMapping{} supported instruction schemes.

\subsection{Prediction Accuracy -- Port Mapping}
\label{ssec:eval:accuracy}

We evaluate our Zen+ port mapping quantitatively by comparing its throughput prediction accuracy against \pmevo~\citep{ritter2020pmevo} and \palmed~\citep{derumigny22palmed}.\footnote{See \cref{sec:relwork} for a conceptual comparison to these approaches.}
As port mappings model only the use of functional units, we focus on instruction sequences whose throughput is not limited by data dependencies.

To predict the throughput of an experiment~$e$ with our mapping, we solve the LP from \cref{sec:background:pm_model} for the number~$t$ of cycles of an optimal execution w.r.t.\ the port mapping.
If this number is faster than the bottleneck of 5 IPC allows, we report an inverse throughput of $5/\abs{e}$ cycles, and~$t$ otherwise.
For \pmevo, we combine the available implementation with the measurement setup used for our case study and infer a new port mapping.
We seed the population of its evolutionary algorithm with 50,000 random port mappings and let it run until evolution converges after ca.\ 59 hours.\footnote{Following the paper, we do not adjust \pmevo's predictions for the IPC bottleneck. Adjusting only causes minor differences in the metrics. }
For \palmed, we use the most recent available model for the Zen architecture.
To keep benchmarking times for \pmevo manageable, we restrict this evaluation to instruction schemes that occur in compiled binaries for the SPEC CPU2017 benchmarks~\citep{spec17} and are covered by our mapping.\footnote{The \palmed model includes data for almost all of the instruction schemes we extracted from \uopsinfo. }
From the resulting \numSchemesInEval{} instruction schemes, we generate 5,000 dependency-free basic blocks, each consisting of five randomly sampled instructions.
We benchmark their throughput in instructions per cycles (IPC) on the Zen+ hardware.

\cref{tab:relaxed_algo:pred_eval} shows the IPC prediction accuracy in terms of mean absolute percentage error (MAPE), Pearson's correlation coefficient (PCC), and Kendall's $\tau_K$ for each tool.
A high PCC indicates a linear correlation of predictions and measurements whereas a high $\tau_K$ implies that sorting the instruction sequences by predicted or measured IPC leads to similar rankings.
Both metrics can range between -1 and 1.

The predictions of \pmevo and \palmed share a similar level of accuracy, with significant correlations but rather high errors of 28--35\%.
Our model is substantially more accurate with an error of 6.7\% and very strong linear and rank correlations.
The heatmaps in \cref{fig:heatmap} quantify each tool's prediction accuracy in more detail.
They group the basic blocks into buckets based on the IPC we observed in the benchmarks and the predictions of each model.
Buckets are displayed darker the more basic blocks they contain; the closer the darker buckets are to the diagonal line (orange) the closer are predictions and observations.
The heatmap for our model, \cref{fig:heatmap:ours}, is notably closer to the diagonal than \pmevo's and \palmed's.

The structure of \pmevo's mapping differs substantially from ours: There, most instructions have only a single kind of \uop in their port usage.
Our explainable approach captures structures of the microarchitecture that \pmevo does not resolve.
As shown in \cref{fig:heatmap:palmed}, \palmed's resource model usually predicts slower executions than what we measure in microbenchmarks.
As \palmed depends on assumptions in its measurement infrastructure, we cannot evaluate whether its model would be more consistent with our throughput measurements if it used our microbenchmarking setup.

\section{Related Work}
\label{sec:relwork}

Aside from \uopsinfo~\citep{abel19uopsinfo}, which the previous sections discuss extensively, two other works in the field of port mapping inference are comparable.
\pmevo \citep{ritter2020pmevo} has even weaker requirements on performance counters than this approach, using only time measurements.
This flexibility comes at the cost of explainability: \pmevo uses an evolutionary algorithm to optimize candidate port mappings such that they accurately model the performance of a fixed set of microbenchmarks.
In contrast to our approach, there is no tangible justification for the entries of port mappings found by \pmevo.
\pmevo uses a heuristically chosen set of benchmarks where complex interactions between instructions might not be represented.

Like our approach, \palmed~\citep{derumigny22palmed} takes inspiration from the \uopsinfo algorithm:
They proceed in two phases, first finding a core mapping for a small instruction set and then inferring models for all other instructions based on the core mapping.
Rather than identifying blocking instructions with a \uop counter, they select basic instructions for their core mapping heuristically based on throughput benchmarks.
Instead of a traditional port mapping, this core mapping (as well as the final result of \palmed) is a \emph{conjunctive mapping} that captures the stress that each instruction puts on various abstract resources of the processor.
\palmed uses (integer) linear programming to construct a set of abstract resources that represent possible bottlenecks in the execution of core mapping instructions.
They further generate for each resource a kernel of basic instructions that saturates the resource, similar to how blocking instructions flood their corresponding set of ports.
\palmed then benchmarks every instruction that is not in the core mapping individually with the saturating kernels and uses a linear program to compute the pressure the instruction puts on the corresponding abstract resources.

Besides port sets, \palmed's resource model inherently represents other potential bottlenecks like the maximal execution rate of the frontend, which our approach needs to treat explicitly.
However, conjunctive mappings are challenging to integrate with existing performance models since the inferred abstract resources have no clear correspondence to documented aspects of the microarchitecture.

Approaches like \ithemal~\citep{mendis19ithemal} and \granite~\citep{sykora2022granite} use machine learning to infer instruction-level throughput models.
Among other factors, they model effects of the utilization of the processor's functional units.
However, as black-box models, they provide less insight into how instructions are executed compared to throughput predictors like \cqa~\citep{rubial14cqa}, \llvmmca~\citep{llvmmca,llvmmca-man}, \osaca~\citep{laukemann18osaca}, \uica~\citep{abel21uica}, or \facile~\citep{abel2023Facile} with an explicit port mapping as we infer it.

\section{Conclusion}
\label{sec:conclusion}

We have shown that per-port \uop counters are not necessary for a \uopsinfo-style port mapping inference algorithm.
If the processor under investigation follows the port mapping model, we can infer the port usage of instructions efficiently.

Our study of AMD's Zen+ microarchitecture indicates that the approach is practical for a large portion of the processor's instructions.
However, there are practical hindrances like throughput bottlenecks in parts of the processor, misdocumented performance counters, and complex micro-coded or non-pipelined instructions.
Nevertheless, we uncovered details of the Zen+ microarchitecture that have, to the best of our knowledge, not been previously documented.
We inferred the first explainable port mapping for over 1,000 instruction schemes on Zen+ that were out of scope for previous work and demonstrated its ability to accurately model performance characteristics of the microarchitecture. 

\appendix
\section{Artifact}

\subsection{Abstract}

This work is accompanied by an artifact that includes our prototype implmentation of the proposed port mapping inference algorithm as well as the data sets and results of the case study in \cref{relaxed:sec:eval}.
In particular, the provided results include human-readable and machine-readable representations of the inferred Zen+ port mapping, the \pmevo and \palmed models used as points of comparison, and the raw data for \cref{fig:heatmap}.
We provide the artifact as a public repository on Github and as an archived virtual machine image bundled with all software dependencies that can be run using Vagrant and VirtualBox.

\subsection{Artifact check-list (meta-information)}

{\small
\begin{itemize}[leftmargin=*]
  \item {\bf Algorithm: } Port Mapping Inference
  \item {\bf Run-time environment: } Python 3, Linux, VirtualBox
  \item {\bf Hardware: } x86-64
  \item {\bf Disk space required:} less than \unitgigabyte{10}
  \item {\bf Publicly available: } \url{https://github.com/cdl-saarland/pmtestbench}
  \item {\bf Code licenses:} MIT
  \item {\bf Archived: } \url{https://zenodo.org/doi/10.5281/zenodo.10794887}
\end{itemize}
}

\subsection{Description}

\subsubsection{How to access}

The artifact can be accessed from the URLs listed in the above check-list.
The source code and data is available via the Github URL whereas the virtual machine image that bundles source code and data with the necessary dependencies is available at the archive URL.

\subsubsection{Hardware dependencies}

The virtual machine image is built for x86-64 processors.

\subsection{Installation}

The virtual machine image comes with installation instructions in the \href{https://github.com/cdl-saarland/pmtestbench/blob/main/vagrant-setup/artifact_usage.md}{artifact\_usage.md} file.
The source code has installation instructions in its \href{https://github.com/cdl-saarland/pmtestbench/blob/main/README.md}{README.md} file.

\subsection{Evaluation and expected results}

The virtual machine image comes with a suggested workflow for evaluating the artifact in the \href{https://github.com/cdl-saarland/pmtestbench/blob/main/vagrant-setup/artifact_usage.md}{artifact\_usage.md} file.
This workflow includes reproducing the heatmaps of \cref{fig:heatmap}.

\bibliographystyle{plain}
\bibliography{references}

\end{document}